\newlength{\singlecolumnfigurewidth}
\newlength{\doublecolumnfigurewidth}
\documentclass[traditabstract]{aa}
\usepackage{natbib}
\bibpunct{(}{)}{;}{a}{}{,}
\usepackage{graphicx}
\usepackage{txfonts}
\begin{document} 

\newcommand{\be}{\begin{equation}}
\newcommand{\ee}{\end{equation}}
\def\msun{M_\odot}
\def\msol{M_\odot}
\def\te{T_{\rm eff}}
\def\roacc{\rho_{\rm acc}}
\def\eacc{e_{\rm acc}}
\def\Ladd{L_{\rm add}}
\def\Aacc{A_{\rm acc}}
\def\vinflow{v_{\rm inflow}}
\def\ur{u_{\rm r}}
\def\sr{S_{\rm r}}
\def\logg{\log g}
\def\lmix{l_{\rm mix}}
\def\lsol{L_\odot}
\def\tconv{\tau_{\rm conv}}
\def\tth{\tau_{\rm th}}
\def\msolyr{M_\odot {\rm yr}^{-1}}
\def\ltsima{$\; \buildrel < \over \sim \;$}
\def\simlt{\lower.5ex\hbox{\ltsima}}
\def\gtsima{$\; \buildrel > \over \sim \;$}
\def\simgt{\lower.5ex\hbox{\gtsima}}

   \title{Multi-dimensional structure of accreting young stars}

   %\subtitle{I. Overviewing the $\kappa$-mechanism}

   \author{C. Geroux\inst{1} \and
          I. Baraffe\inst{1, 3, 2} \and
          M. Viallet\inst{2} \and
          T. Goffrey\inst{1} \and
          J. Pratt\inst{1} \and
                    T. Constantino\inst{1} \and     
          D. Folini\inst{3} \and
          M. V. Popov\inst{3} \and
          R. Walder\inst{3} 
          }

   \institute{University of Exeter, Physics and Astronomy, EX4 4QL Exeter, UK,
              \email{geroux@astro.ex.ac.uk} \and
             Max-Planck-Institut f{\"u}r Astrophysik, Karl Schwarzschild Strasse 1, 85741 Garching, Germany \and
             {\'E}cole Normale Sup{\'e}rieure de Lyon, CRAL (UMR CNRS 5574), Universit{\'e} de Lyon 1, 69007 Lyon, France
             }

   \date{}

% \abstract{}{}{}{}{} 
% 5 {} token are mandatory
 
  \abstract
   {This work is the  first attempt to describe the multi-dimensional structure of accreting young stars based on fully compressible time implicit multi-dimensional hydrodynamics simulations. One major motivation is to analyse the validity of accretion treatment used in previous 1D stellar evolution studies. 
We analyse the effect of accretion on the structure of a realistic stellar model of the young Sun. Our work is inspired by the numerical work of Kley \& Lin (1996, ApJ, 461, 933) devoted to the structure of the boundary layer in accretion disks, which provides the outer boundary conditions for our simulations. We analyse the redistribution of accreted material with a range of values of specific entropy relative to the bulk specific entropy of the material in the accreting object's convective envelope. Low specific entropy accreted material characterises the so-called ``cold" accretion process, whereas high specific entropy is relevant to ``hot" accretion. A primary goal is to understand whether and how accreted energy deposited onto a stellar surface is redistributed in the interior. This study focusses on the high accretion rates characteristic of FU Ori systems. We find that the highest entropy cases produce a distinctive behaviour in the mass redistribution, rms velocities, and enthalpy flux in the convective envelope. This change in behaviour is characterised by the formation of a hot layer on the surface of the accreting object, which tends to suppress convection in the envelope. We analyse the long-term effect of such a hot buffer zone on the structure and evolution of the accreting object with 1D stellar evolution calculations. We study the  relevance of  the assumption of redistribution of accreted energy into the stellar interior used in the literature. We compare results obtained with the latter treatment and those obtained with a more physical accretion boundary condition based on the formation of a hot surface layer suggested by present multi-dimensional simulations. 
One conclusion is that, for a given amount of accreted energy transferred to the accreting object, a treatment assuming accretion energy redistribution throughout the stellar interior could significantly overestimate the effects on the stellar structure and, in particular, on the resulting expansion. 
%One conclusion is that a treatment assuming accretion energy redistribution throughout the stellar interior can mimic the effect of a more physical accretion boundary condition based on the formation of a hot surface layer, which is suggested by the present multi-dimensional simulations. This is an important result because it does not invalidate previous studies based on the former simplistic treatment for hot accretion.

 % results heading (mandatory)
   %We find that depending on the entropy of the accreted material relative to the average entropy in the convective region two main behaviours occur: 1) for entropies lower than $\sim 40$\% larger than the average convective entropy, the accreted material is mixed in via convection, 2) for entropies higher than $\sim 40$\% larger than the average convective entropy, the accreted material remains very near the surface as convection close to the surface has ceased and does not mix in the newly accreted material.
   }


   \maketitle
%
%________________________________________________________________

\section{Introduction}

\label{introduction}

Accretion is an important process relevant to various fields in astrophysics, from star formation to the study of compact binaries and supernovae type Ia. Continuous theoretical and numerical efforts have been devoted to understanding the interaction between the inner edge of an accretion disk and the stellar surface. Analysis of this transition region is often based on boundary layer models that characterise this narrow region where the accreted matter leaving the disk decelerates to join the surface of the star, white dwarf or neutron star \citep[see][and references therein]{Pringle-1981}. Other approaches to studying the spread of accreted matter over the stellar surface are based on the so-called ``spreading layer'' model, following analytical models applied to neutron stars or white dwarfs \cite[e.g.][]{Inogamov-1999, Piro-2004}. Several authors have attacked the problem with multi-dimensional numerical methods \cite[see e.g.][]{Kley-1987, Kley-1996, Balsara-2009, Hertfelder-2013}. The numerical efforts from \cite{Kley-1987} and \cite{Kley-1996}, which were devoted to the accretion process on protostars, demonstrated the importance of meridional flow development over the surface of the accreting object, showing in the case of high accretion rates (FU Ori type stars) how matter  accreted from a disk can spread over the poles and completely engulf the protostar.

Complementary to these works, there are also many studies devoted to the effect of accretion on the structure and evolution of  objects, using either polytropic approaches \cite[e.g.][]{Hartmann-1997} or simulations based on one dimensional (1D) stellar evolution codes \cite[e.g.][]{Mercer-1984, Prialnik-1985, Shaviv-1988, Siess-1997, Baraffe-2009, Hosokawa-2010}. Despite many efforts, great uncertainties still remain regarding the amount of accretion luminosity imparted to the accreting object or radiated away, how the accreted material is redistributed over the stellar surface and the depth it reaches. Various treatments have been used, based either on modifications of the outer boundary conditions at the stellar surface \cite[e.g.][]{Mercer-1984, Shaviv-1988, Palla-1992, Hosokawa-2010} or assuming that some amount of thermal energy released by accretion, and characterised by a part of the accretion luminosity, is added to the stellar interior  \cite[e.g.][]{Hartmann-1997, Siess-1997, Baraffe-2009}. In the latter case, accretion is assumed to proceed through a disk
with a part $\epsilon$ ($\le 1/2$) of the gravitational potential energy  going into the internal energy of the gas in the boundary layer. As material is incorporated into the star from the boundary layer, a part $\alpha$ of the energy of the gas is absorbed by the protostar and a part $(1-\alpha)$  is radiated away. This leads to a rate where energy is added to the accreting protostar
\begin{equation}
\Ladd =\alpha\epsilon\frac{GM\dot{M}}{R},
\label{eq:acc-eng}
\end{equation}
where $M$ is the mass of the accreting protostar, $\dot{M}$ the accretion rate, and $R$  the stellar radius. The parameter $\alpha$ is free and characterises the accretion process, with $\alpha=0$ for  ``cold" accretion and $\alpha>0$ for ``hot" accretion \citep{Baraffe-2012}.

The importance of understanding the effects of accretion on stellar structure and  evolution  was again highlighted  in a series of works from one of the present authors \citep{Baraffe-2009,Baraffe-2010, Baraffe-2012} within the context of early stages of evolution of young low mass stars and brown dwarfs.  Early accretion history may affect the properties of objects (luminosity, radius, and effective temperature) in star formation regions and young clusters even after several Myr when the accretion process has ceased. This scenario can provide  an explanation for a well known observed feature, namely the luminosity spread in Hertzsprung-Russel diagrams of star-forming regions and young clusters. \cite{Baraffe-2012} proposed a so-called ``hybrid" accretion scenario,  wherein the amount of accretion energy  absorbed by the protostar depends on the accretion rate.  
This hybrid scenario  led to a proposed unified picture linking the observed luminosity spread in young clusters and FU Ori eruptions.

The work by \citeauthor{Baraffe-2009} could have important consequences for our understanding of star formation regions and young clusters. But the conclusions rely on a very simplistic approach to accretion in their stellar evolution calculations where  the accreted material and energy are added instantaneously and uniformly throughout the stellar model. A more sophisticated model to describe the redistribution of accreted mass and energy within young stars has been developed by \cite{Siess-1996, Siess-1997}. Their approach is based on earlier models by \cite{Kippenhahn-1978,Kutter-1987} developed for the study of accreting white dwarfs  within the context of nova outburts. The basic idea is that the accreting material possesses angular momentum and its accretion onto a slowly (or non-) rotating stellar surface causes shear instabilities that induce mixing of material and transport the accreted angular momentum. The redistribution of mass, energy and angular momentum can be characterised by the same ``penetration function", which depends among other parameters, on the Richardson number $R_{\rm i}$ defined as the ratio of potential energy of buoyant force to kinetic energy of turbulence.
%characterising  stability against shear instabilities. 
The Siess et al. formalism is rather complex and it is difficult to gauge the added physical correctness achieved by this sophisticated model. Interestingly, while there have been several attempts to tackle the problem of boundary layers of accretion disks with multi-dimensional approaches, as previously mentioned, no multi-dimensional (hereafter multi-D) numerical study has ever been devoted to  the effect of accretion on the structure of accreting objects. 

This work is the very first attempt to describe with two dimensional (2D) hydrodynamic simulations the redistribution of accreted material and energy in accreting young stars. One motivation is to test some of the main assumptions used in the scenario developed by \cite{Baraffe-2009, Baraffe-2012}, which heavily relies on the hypothesis of ``cold" and ``hot" accretion. In particular, the success of explaining various observations depends on the possibility of efficiently redistributing a given amount of the accreted energy within the stellar interior. More generally, the aim of such multi-dimensional simulations is to provide a clearer picture of the accretion process onto young stars and provide  physical support to simple treatments of accretion used in stellar evolution codes for the wide range of astrophysical applications above-mentioned. 

The hydrodynamical simulations were computed using our fully compressible time implicit code MUSIC (Multidimensional Stellar Implicit Code), initially developed by \cite{Viallet-2011, Viallet-2013} and recently improved with the implementation of a Jacobian-free Newton-Krylov time integration method, which significantly improves the performance of the code \citep{Viallet-2016}. This first numerical study devoted to accretion is inspired by the numerical work of \cite{Kley-1996} concerning the structure of the boundary layer in accretion disks and which provides  the outer boundary conditions for our simulations. One of our primary goals is to understand whether and how accreted energy deposited onto a stellar surface can be redistributed in the interior. This first study focusses on high accretion rates characteristic of  FU Ori systems and relevant to the  
\cite{Baraffe-2009, Baraffe-2012} analysis. 

Our paper is organised as follows. In  \S~\ref{model} we describe the stellar model used as a ``laboratory" for our numerical experiments and the implementation of the accretion process in MUSIC. We analyse in \S~\ref{structure}  the effect of accretion on the stellar structure and in \S~\ref{distribution} the redistribution of accreted material assuming it has different values of specific entropy relative to the bulk specific entropy of the accreting object. Low specific entropy of the accreted material characterises the so-called ``cold" accretion process whereas high specific entropy is relevant for the ``hot" accretion case. The main findings resulting from the 2D calculations are transposed in our 1D stellar evolution code in \S~\ref{1D} in order to understand the impact of our results on the evolution of accreting young stars. Finally, we discuss and conclude on the limitations  and implications of these first multi-D results in \S~\ref{discussion}. 

\section{Model of Stellar Accretion}
\label{model}

\subsection{Initial stellar model}
\label{initial}

Our main goal is to study the process of accretion on young stellar objects, which are essentially fully convective.
We  adopt as a stellar ``laboratory" the young Sun model already used in \cite{Viallet-2013, Viallet-2016} for the development and benchmarking of MUSIC. The main characteristics of MUSIC are briefly recalled below. It is a fully time implicit code solving the compressible hydrodynamic equations in spherical geometry using a finite volume method. Radiation transport is modelled with the diffusion approximation and realistic stellar equation of state and opacities are included. The code is continuously being developed and benchmarked and all numerical details are provided in the \citeauthor{Viallet-2016} series of papers. Thanks to the time implicit solver and the spherical coordinate system, MUSIC is particularly well suited for the description of convection in stellar interiors. A systematic study devoted to the sensitivity of convection properties to numerical treatments, including the resolution and extension of the spatial grid and the effects of boundary conditions, is presented elsewhere (Pratt et al. 2016, in prep.). 
%We wish to study how material is distributed within an accreting star, to do so we simulate the convective flow using the MUSIC code \citep{Viallet-2011} which was developed with the specific goal of simulating convection within stellar interiors. It is a fully time implicit code solving the compressible hydrodynamic equations in spherical geometry using a finite volume method. Radiation transport is modelled with the diffusion approximation and realistic stellar equation of state and opacities are included.
The young solar model  was generated using the Lyon stellar evolution code \citep{Baraffe-1991, Chabrier-1997,Baraffe-1998}.
It is a 1\,$\msol$ star of a few Myr, with radius $R \sim 3\,R_\odot$, effective temperature $\te \sim$ 4100 K and luminosity $L \sim 2.3\,L_\odot$, still contracting toward the Main-Sequence. Those properties are very similar to that of the central protostar adopted by \cite{Kley-1996}. The model  has developed a radiative core that covers $\sim$ 40\% of the stellar radius and a fully convective envelope covering the remaining 60\% of the star. Note that energy generation from deuterium burning is not important for this particular model and does not need to be accounted for in the multi-D simulations.

\begin{figure}[h]%1
\centering
\includegraphics[width=\singlecolumnfigurewidth]{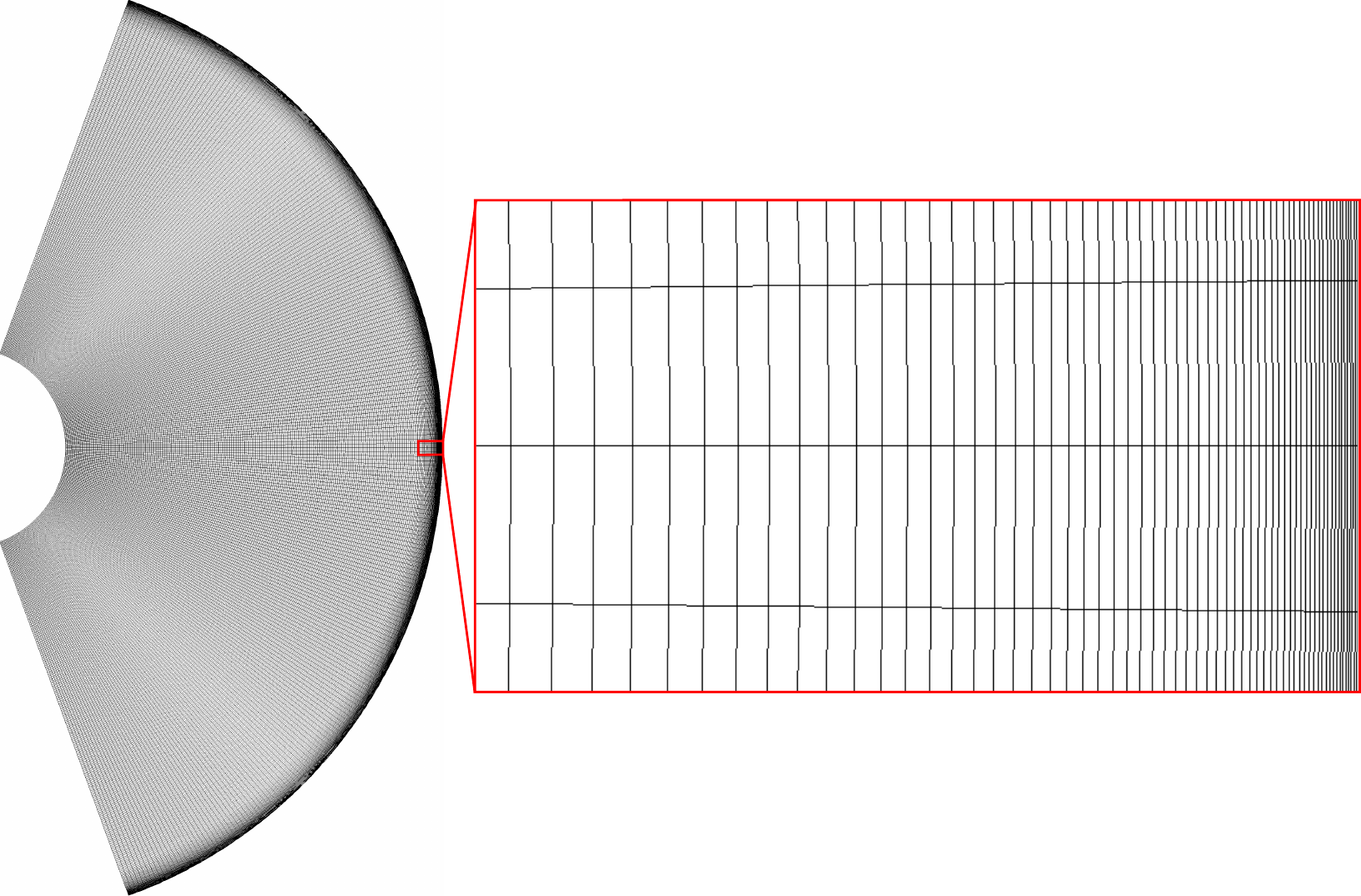}
  \caption{Visualisation of the computational mesh. On the right is a zoom of the near surface mesh showing the geometrically decreasing radial spacing. Note that the  changes in grid-lines are a result of the rasterisation process to generate the figure and not indicative of a  change in actual grid geometry.}
     \label{fig:grid}
\end{figure}

The initial model for MUSIC is spherically symmetric with resolution in ($r$, $\theta$) of $320 \times 256$ and uniform chemical composition and is created from the 1D model. Special attention is paid to the radial grid close to the surface. In the 1D model, the surface is defined using the Eddington approximation, with the surface corresponding to the photosphere at an optical depth $\tau$ = 2/3 and a surface temperature $T_{\rm surf}$ equal to the effective temperature $\te$ \cite[see e.g.][]{Chabrier-1997}.  The small scale surface convection resulting from the rapidly decreasing pressure scale height near the surface of a star as well as steep surface radial gradients require a smaller radial spacing near the surface. We thus design the radial grid for the 2D simulations based on the following procedure. An initial radial spacing $\Delta r_0$ at the surface of the stellar model is defined. Then, for each new radial cell, $i$, a new $\Delta r_i=\Delta r_{i-1}(1+\epsilon)$ is calculated until a fractional stellar radius $r/R=0.94$ is reached (with $r/R$=1.0 defining the surface). In the present work, we use $\epsilon = 0.05$, yielding 64 shells in this region.
Below $r/R=0.94$ we use a fixed $\Delta r$ such that 256 cells are evenly spaced between $r/R=0.94$ and $r/R=0.2$. Figure~\ref{fig:grid} shows a visualisation of the resulting grid, with a zoom of the outer geometrically decreasing radial spacing. 
%The smaller radial spacing near the surface allows the simulation to better resolve the small scale surface convection resulting from rapidly decreasing pressure scale height near the surface as well as steep radial gradients found at the surface. 
In the angular direction we use 256 evenly spaced divisions. In the present case, we model a region covering $\theta=20^\circ-160^\circ$, with $\theta$ measured clockwise from the vertical axis.

Initially the 2D model has zero radial and tangential velocities. The radial grid extends down to the middle of the  stellar radiative core. In 2D, MUSIC uses as independent variables the density $\rho$, the specific internal energy $e$, the radial velocity $u_{\rm r}$ and the tangential velocity $u_{\rm t}$\footnote{Conserved variables are however used to solve the conservation equations \cite[see][for details]{Viallet-2016}.}.
Periodic conditions in the tangential direction are used. The boundary conditions for velocities at the top and the bottom of the domain are based on reflective boundary conditions for the radial velocity and stress-free conditions for the tangential velocity \citep[see][]{Viallet-2011}. 
At the bottom of the domain,  a constant energy flux, derived from the 1D initial model, is imposed and linear extrapolations as a function of radius in density and energy are used to set the values in the ghost cells. 
At the top of the domain, i.e. the ``surface", the energy flux is given by $F=\sigma T^4$, with $\sigma$ the Stefan-Boltzmann constant and $T$ the temperature of the cells in the top layer of the domain, in order to remain consistent with the 1D initial model. During the relaxation process of the initial model, the surface value of the density $\rho_{\rm surf}$ is forced to be equal to that of the 1D initial model. The values of $\rho$ and $e$ in the ghost cells are the same as in the cells of the top layer. Note that values of density and energy in the ghost cells at the top and bottom of the domain are mainly used for higher order reconstruction of interface values.
Uncertainties and limitations due to the boundary conditions used for present study are discussed in \S~\ref{discussion}.

After a short time ($< 10^6$\,s) convection develops. The initial model requires some time  to relax to a nearly steady state as indicated by the steady evolution of the total kinetic energy of the model with time. After this relaxation time, 
 the model has a well developed convective region from a fractional radius $r/R= 0.41$ to the surface. %Below this region is a convectively stable radiative region. 
 For this particular initial model, the relaxation time to reach steady-state convection is $\sim$ $2 \times 10^7$\,s, which corresponds to $\sim$ 50 convective turnover timescales (see \S \ref{structure}).
Note that this relaxation time depends on the initial model, its radial extension and the boundary conditions, as analysed in a follow-up paper (Pratt et al. 2016).
This relaxed model is the starting point for exploring the accretion process, i.e. the accretion boundary condition is only applied after this period of relaxation. 
%The simulation times we refer to in the following include this relaxation time.  For example at a simulation time of $3.5\times 10^7$\,s, the model has been accreting for $1.5\times 10^7$\,s. 

%\begin{figure}[h]%1b
%\centering
%\includegraphics[width=44mm]{./figures/f2.pdf}
%  \caption{Geometry of accretion boundary condition. The solid outlined pie slice represents the computational domain, the dashed lines represent the opening angle of the accretion disk with half angle, $\Delta \theta$ centered on the stellar equator. The outer surface boundary condition of the stellar model within the region of the accretion disk is given a fixed density, $\roacc$, and fixed specific internal energy, $\eacc$. An inflow condition is specified with a radial inflow speed of $\vinflow$.}
%     \label{fig:acc-geo}
%\end{figure}

\begin{table*}
\caption{Summary of simulation parameters and results. $\delta s_{\rm acc}$ is the difference between the  specific entropy of the accreted material and the average specific entropy in the convective zone. $\eacc$ is the specific internal energy of the accreted material. $\tconv$ is the convective turnover time scale. $T_{\rm surf}$ and $\rho_{\rm surf}$ are the volume-weighted average of the surface temperature and density, respectively (those  quantities  are calculated after the accretion boundary condition is applied and a quasi-steady state equilibrium is reached). $L$ is the total luminosity, calculated from the Stefan-Boltzmann law and assuming $T_{\rm surf}$ is representative of the stellar effective temperature.
%, and $\Delta T_{\rm surf.}$ and $\Delta\rho_{\rm surf}$ are the total variation of temperature and density at the surface shell over all $\theta$-zones.
}
\label{table:sim-sum}
\centering                          % used for centering table
\begin{tabular}{c c c c c c c c }        % centered columns (4 columns)
\hline\hline                 % inserts double horizontal lines
model & $\delta s_{\rm acc}$ & $\eacc\ 10^{12}\ {\rm erg\ g^{-1}}$ & $\tau_{\rm conv}10^5$ s & $T_{\rm surf}\ {\rm K}$ & $L/{\rm L}_{\odot}$ & $\rho_{\rm surf}\ 10^{-7}{\rm g\ cm^{-3}}$ \\   % table heading 
\hline                       
mdot0 &    -- &   -- & 3.82 &  3880 &   1.89 & 3.9  \\
e1    & -30\% & 2.78 & 2.97 &  4930 &   4.94 & 2.3  \\
e2    & -20\% & 4.74 & 2.89 &  5580 &    8.10 & 2.2  \\
e3    & -10\% & 6.68 & 3.10 &  6080 &    11.4  & 2.4  \\
e4    &   0\% & 8.66 & 2.81 &  6460 & 14.5  & 2.3  \\
e5    & +10\% & 10.7 & 2.95 &  6810 &   18.1  & 2.0  \\
e6    & +20\% & 13.3 & 3.24 &  7190 &   22.3  & 2.0  \\
e7    & +30\% & 19.1 & 3.36 &  7890 &  32.3  & 2.1  \\
e8    & +40\% & 30.9 & 4.73 &  8890 &   52.2  & 1.9  \\
e9    & +50\% & 66.3 & 6.14 & 10790 &   113    & 2.4  \\
\hline                                   %inserts single line
\end{tabular}
\end{table*}

\subsection{Multi-dimensional treatment of accretion}
\label{accretion}

Accretion is implemented by applying a simple inflow boundary condition to the surface of the relaxed convective model. This is a very simplistic treatment which is far from treating the process of accretion with a fundamental physical approach. But this is a reasonable working hypothesis for our numerical experiments, given our first goal, which is to study with multi-D configurations the structure and evolution of accreting stars and to compare  that to 1D approaches.
%Figure~\ref{fig:acc-geo} shows the configuration of our accretion boundary condition. The pie slice with the solid outline represents the portion of the star we are modelling ($\theta=20^\circ-160^\circ$). 
%In our case we modelled a region covering $\theta=20-160^\circ$, with $\theta$ measured clockwise from vertical. 
%The smaller pie slice outlined with dashed lines represents the region over which we apply our accretion boundary condition and has a half opening angle of $\Delta \theta$. 
The boundary values which must be specified in MUSIC for our accretion boundary condition are the inflow speed, $\vinflow$,  density, $\roacc$, and  specific internal energy, $\eacc$, of the accreted material. These values are set and held constant in the outer ghost cells of our simulation.  $\roacc$ is determined from $\vinflow$ and the mass accretion rate $\dot{M}$  such that 
\be
$$\roacc=\dot{M}/(\Aacc \vinflow), $$
\ee
where $\Aacc$
%=4\pi R^2 \sin(\Delta\theta)$ 
is the area over which the accretion boundary condition is applied. 
%This area covers $4\pi$ degrees in the $\phi$-dimension. 
%Outside the accretion area, we use the same outer boundary conditions as in \S \ref{initial}.
%we use boundary conditions of a constant extrapolation in $\rho$, $e$, $v_\theta$, and reflective in $v_r$. The values of the energy in the ghost cells are mainly used for higher order reconstruction of interface values, as the energy flux is set at the surface using the Stefan-Boltzman law $F=sigma T^4$, where $T$ is the temperature of the last simulated cell. In our original 1D model the last temperature included is that of the photosphere, and thus equal to the effective temperature of the star.

The work by \cite{Kley-1996} is used as a guide to set the values for $\Aacc$ and $\vinflow$. 
As mentioned in \S \ref{introduction}, we focus in this paper on high accretion rates and adopt $\dot{M}=10^{-4}   
 \msolyr$, which corresponds to model 2 of \cite{Kley-1996}. For such high accretion rates, the \cite{Kley-1996} simulations suggest that the boundary layer covers the entire surface of the star. This picture is also consistent with other models based on the idea of a ``spreading layer"  where the infalling gas does not decelerate in the midplane of the disk but rather across the whole surface of the star \citep[e.g.][]{Inogamov-1999}. 
 We thus assume $\Aacc$ to be equal to the surface area of the portion of the star we are modelling, i.e. $\Aacc
=4\pi R^2 \sin(\Delta\theta)$ with $\Delta\theta = 70 ^\circ$.
% We thus  set $\Delta\theta$ to cover the entire domain of our simulation. 
%We have chosen a relatively high mass accretion rate of $\dot{M}=10^{-4} \msolyr$  and from \citeauthor{Kley-1996}'s model2, which had the same accretion rate, we see that the boundary layer has covered the surface of the star, thus we set $\Delta\theta$ to cover the entire domain of our simulation. 
As an estimate for the inflow speed we choose one tenth of the sound speed at the surface of the non-accreting model, or $\vinflow=6.5\times 10^4$ cm s$^{-1}$ because \cite{Kley-1996} predict infall velocities of about this order of magnitude at the surface of the star. The above values yield an accretion boundary condition for the density of $\roacc$ $ \sim 2.7 \times 10^{-7}$ g cm$^{-3}$. 
%We initially considered much lower accretion rates of $\dot{M}=10^{-6}$ M$_\odot$ yr$^{-1}$ however, over the time in which we were able to simulate we saw very little change in convective flow patterns and thus in this paper we focus mainly on the higher, $\dot{M}=10^{-4}$ M$_\odot$ yr$^{-1}$, accretion rates.

The final boundary condition that needs to be specified is the specific internal energy of the accreted material, $\eacc$.  Determining suitable values of $\eacc$ is a radiation hydrodynamics problem that requires a boundary 
(or spreading) layer model including the description of the accretion disk. This is a problem disconnected from our study and which cannot be addressed with our MUSIC code.
%It is difficult to say exactly what this value should be, however, 
We can thus only explore a range of values for $\eacc$. Expectedly, if the accreted material is ``cold" with a relatively low entropy, compared to the bulk entropy in the convective envelope, it will sink. 
Conversely hotter material with larger entropy will remain near the surface. With a given equation of state, the specific entropy of the material can be related to the specific energy and density of that material. 
The convective envelope in the young Sun is essentially isentropic with a specific entropy of $s=3.14\times10^9$~erg~K$^{-1}$~g$^{-1}$. We explore a range of values for $\eacc$, so that the specific entropy of the accreted material $s_{\rm acc}$ covers values 30\% lower to 50\% higher than the specific entropy in the convective region, with the intention that material with lower $s_{\rm acc}$  will mix into the convective region, while that with  higher $s_{\rm acc}$ will remain at the surface. 

In order to link the arbitrary variation of $\eacc$ to stellar evolution studies of accreting objects, we note that
the value of $\eacc$ can  directly be related to  $\alpha$ in Eq. (\ref{eq:acc-eng}) by multiplying $\eacc$ by $\dot{M}$ to get a rate of energy accretion. The lowest value of $\eacc$ adopted in this work, equivalent to 30\% lower entropy, corresponds to $\alpha \sim 0.01$, and the highest $\eacc$, equivalent to 50\% higher entropy, corresponds to $\alpha \sim 0.2$. This range of values for $\alpha$ is consistent with those adopted in \cite{Baraffe-2012}. 
Table~\ref{table:sim-sum} lists the various models we have computed, along with some relevant parameters and values of the simulation. We cover many cases in order to be able to find an eventual transition in the behaviour of accretion. The total simulation time for all cases is $3\times 10^7$ s.
For all our simulations, except the highest accreted entropy case (case e9 in Table~\ref{table:sim-sum}), the timestep is characterised by a CFL number between typically $\sim 10$ and $\sim 80$, which is essentially set  by the radiative CFL number \cite[see][for definitions]{Viallet-2011}. The hydrodynamic CFL number, CFL$_{\rm hydro}$, varies from $\sim$ 1.5 to $\sim$ 10 depending on the simulation. For the e9 case, the timestep reduces to an average CFL number $\sim$ 2.5. Note that, for the sake of accuracy and to help convergency, 
the timestep is also constrained  by a condition that limits the relative change in energy and density from one time step to the next to less than 1\%. Reducing this constraint could allow larger CFL numbers. 
%The quantities $\eacc$ and $s_{\rm acc}$ we have just mentioned, the remaining values in Tabel~\ref{table:sim-sum} will be discussed in the following section.

\section{Effects of accretion on the stellar structure}
\label{structure}

\begin{figure}[h]%3
\centering
\includegraphics[width=8cm]{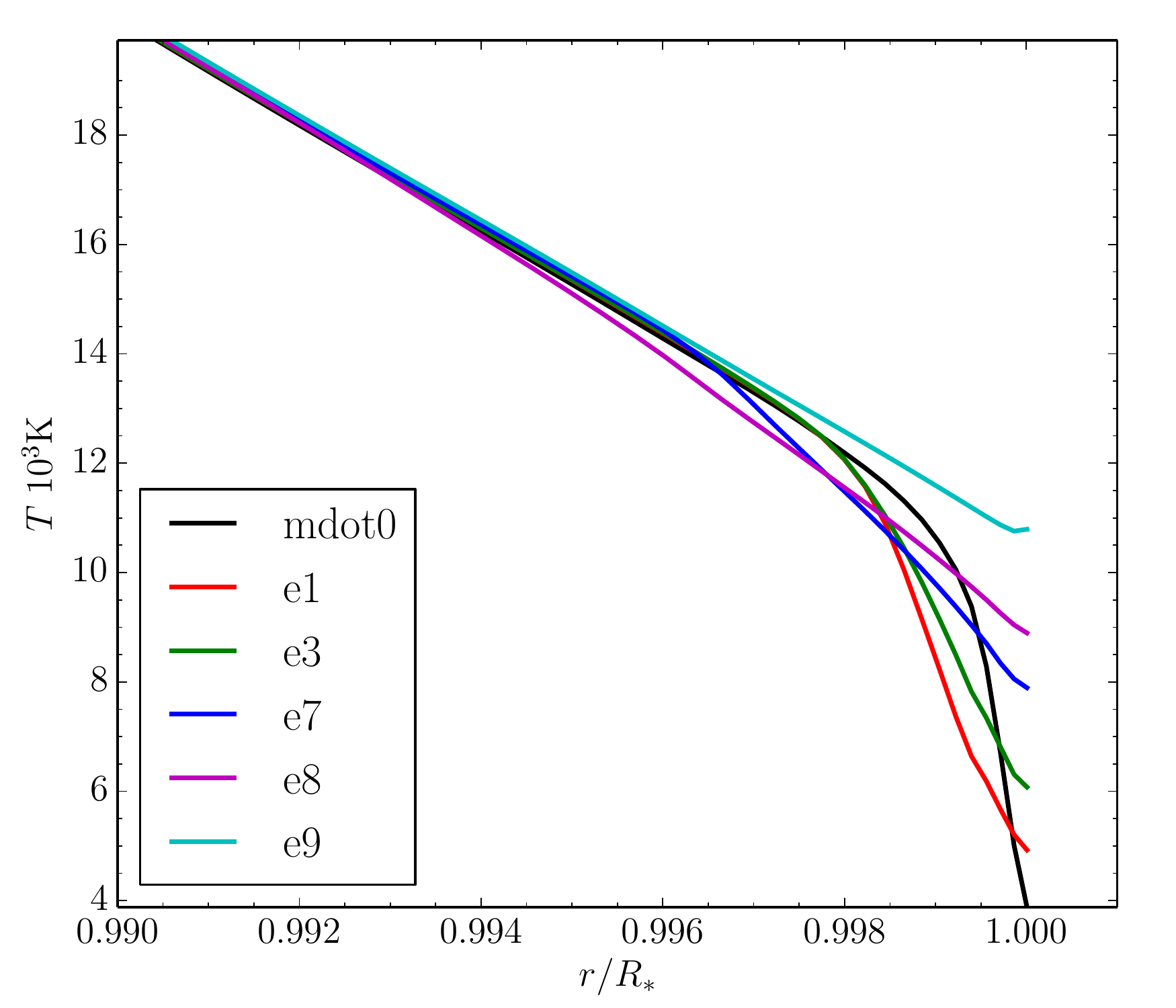}
  \caption{Volume-averaged temperature  in the outer 1\% of the stellar radius for the non-accreting model (mdot0) and for the accreting models (cases e1 to e9) at time  $t=3\times 10^7$~s. At this stage $1.9\times 10^{29}$~g has been accreted ($3\times 10^7$~s of accretion at an accretion rate of $1\times 10^{-4}~M_\odot$~yr$^{-1}$). Curves labelled e1-e9 correspond to the accreting calculations with varying $\eacc$ (see Table~\ref{table:sim-sum} for the specific internal energies of the accreted material).}
     \label{fig:temp-struct}
\end{figure}

\begin{figure}[h]%4
\centering
\includegraphics[width=8cm]{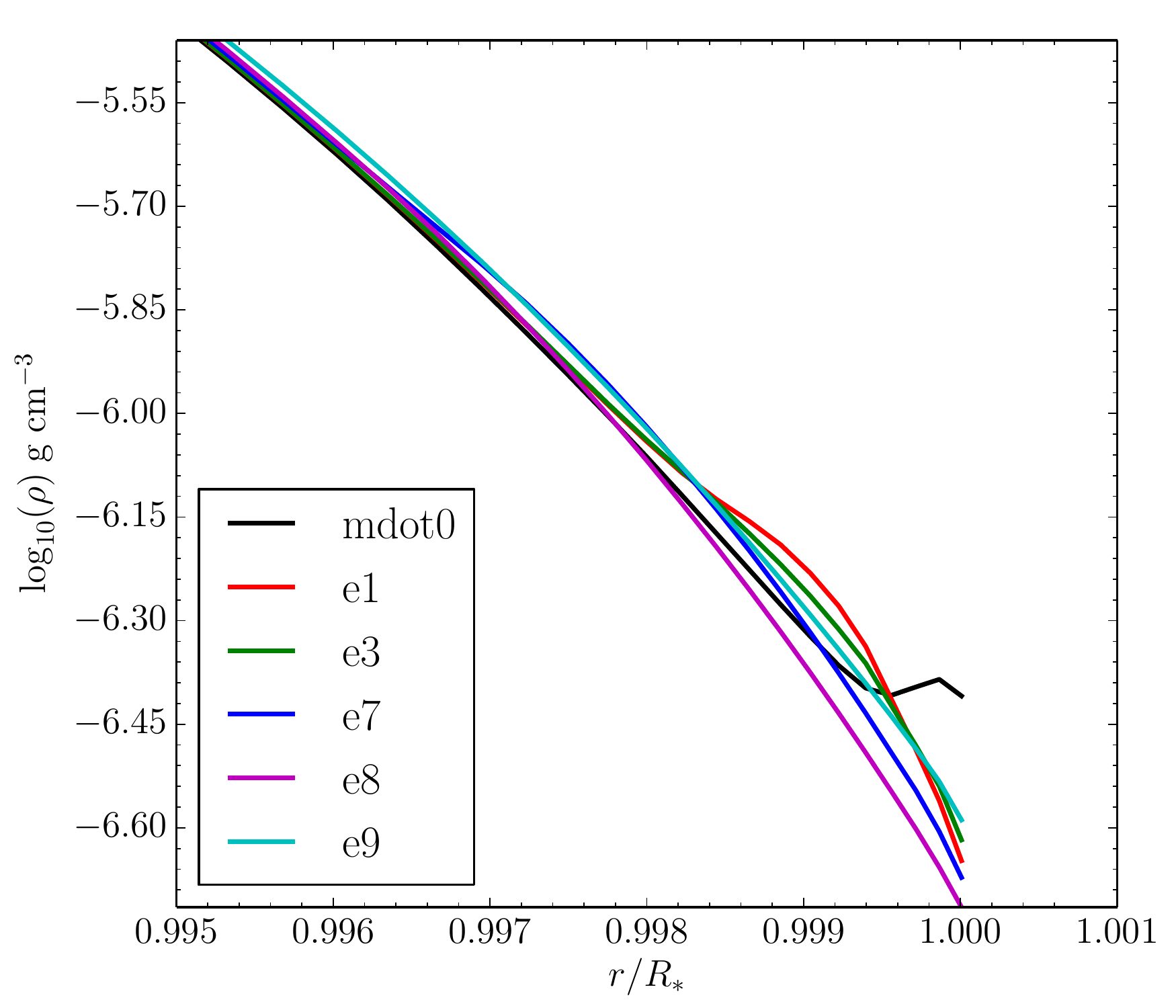}
  \caption{Volume-averaged density  in the outer 0.5\% of the stellar radius for the models shown in Figure~\ref{fig:temp-struct}.}
\label{fig:density-struct}
\end{figure}

When the accretion boundary condition is applied to the relaxed non-accreting model, the surface structure of the model adjusts very rapidly to a new equilibrium configuration, taking between $2\times 10^5$ to $2\times 10^6$\,s for all significant adjustments, depending on the accretion energy, with higher accretion energies taking less time to adjust. The rate of adjustment depends on the depth of the region in the model, with deeper regions taking longer to adjust than shallower regions (see \S \ref{1D}). Going deeper into the stellar interior the magnitude of the changes due to accretion decreases.  Figures~\ref{fig:temp-struct}~and~\ref{fig:density-struct} show, respectively, the temperature and density structures close to the surface of accreting and non-accreting models. As one would expect,  the surface temperature is highest  for the hottest accretion case, e9. The effect on the density structure is more complex, as it combines the mixed effects due to temperature change, mass addition and compression work from the infalling material.  
At the very surface, all accreting models have a slightly lower surface density (see Table \ref{table:sim-sum}) relative to the non-accreting model,  in part due to the imposed accretion boundary condition on the density $\rho_{\rm acc}$. Slightly deeper, the density profiles readjust differently for models with different
accretion energy input, depending on the heating efficiency and the mass redistribution of the accreted matter (see \S \ref{distribution}). 
%, but has increased slightly deeper just below the hydrogen ionization zone. 
After the relatively short initial readjustment the average radial structure remains relatively steady.

\begin{figure}[h]%5
\centering
\includegraphics[width=\singlecolumnfigurewidth]{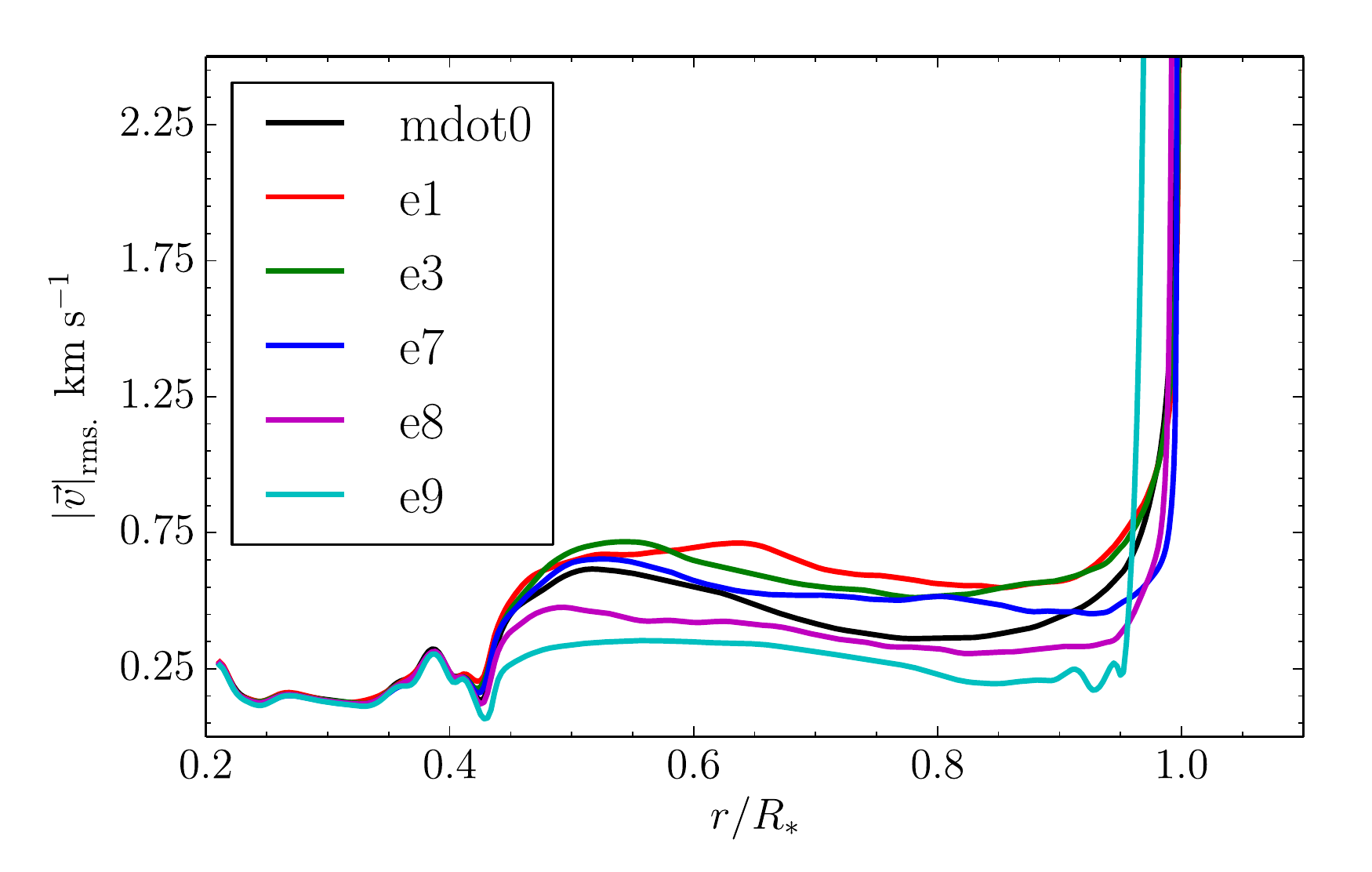}
\caption{Profile of the  rms velocities volume-weighted averaged over $\theta$ and time-averaged over $10^7$ s  for the models shown in Figure~\ref{fig:temp-struct}. 
%The time average covers  $10^7$~s ($\sim 20 \times \tconv$) of simulation time for the models shown in Figure~\ref{fig:temp-struct}. %The upper panel shows the velocity magnitude, the middle panel shows the $r$-component, and the bottom panel shows the $\theta$-component of the velocity ***I DO NOT WANT TO SHOW MIDDLE AND LOWER PANEL***. 
%Curve labelling is similar to Figure~\ref{fig:temp-struct}.
}
\label{fig:vrms-1e-4}
\end{figure}

%\begin{figure}[h]%6
%\centering
%\includegraphics[width=\singlecolumnfigurewidth]{./figures/f6.pdf}
%  \caption{***DELETE FIGURE***The difference between the actual temperature gradient and the adiabatic temperature gradient. Positive values indicate instability to convection while negative values indicate stability to convection. Models mdot0-e7 all show strong instability to convection while e9 shows a mild stability to convection at fractional radii between 0.996-0.999.}
%\label{fig:super-ad}
%\end{figure}

\begin{figure}[h]%6
\centering
%\vspace{-1cm}
\includegraphics[width=\singlecolumnfigurewidth]{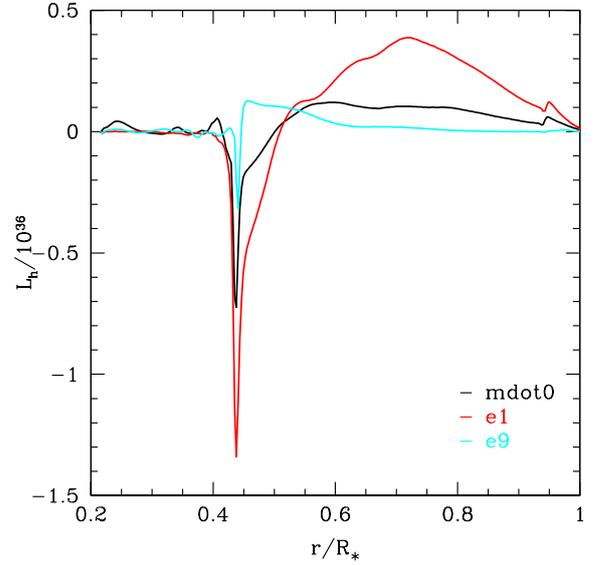}
\vspace{-2cm}
  \caption{Radial profiles of the time-averaged enthalpy luminosity (in erg s$^{-1}$) in the non-accreting model (black line) and accreting cases e1 (red line) and e9 (cyan line).}
\label{fig:lenthalpy}
\end{figure}

%From only the variation in density and temperature with $\theta$ given in Table~\ref{table:sim-sum} as $\Delta \rho_{\rm surf.}$ and $\Delta T_{\rm surf.}$ one might get the hint that convection strength is dissipating in the higher energy accretion calculations relative to the non-accreting simulation. 

The increase in  the temperature of the surface layers as $\eacc$ increases (see Figure~\ref{fig:temp-struct}) affects the energy transport and the properties of the convection in the envelope. We first  analyse the convective velocities for all considered models. These can be characterised by the averaged root-mean-square (rms) velocities $v_{\rm rms}$ in the convective region. 
%To see how the accretion is affecting the convective velocities 
Figure~\ref{fig:vrms-1e-4} shows the  rms velocities volume-weighted averaged in the $\theta$-coordinate and time-averaged over $10^{7}$\,s. 
This time interval is sufficient for calculating time-averaged quantities because it equates to between $\sim$ 17 and $\sim$ 36 convective turnovers, depending on the models (see Table ~\ref{table:sim-sum}). 
%Estimating a convective turn over time scale shows that averaging over this period of time is relevant. 
The convective turnover time scale, $\tconv$, can be defined by:
\begin{equation}
\tconv=\left\langle \frac{ \int_{ r_{\rm core} }^{ r_{\rm surf}} \langle\tconv\rangle_\theta \, dr }{ \int_{r_{\rm core}}^{r_{\rm surf.}} dr}\right\rangle_{t}
\end{equation}
where
\begin{equation}
\langle\tconv\rangle_\theta=\langle H_{\rm p}\rangle_\theta / v_{\rm rms}.
\end{equation}
Here, $r_{\rm core}$ and $r_{\rm surf}$ are the radius at the bottom and top of the convective zone, respectively, and  $H_{\rm p}$ is the pressure scale height.
The angle brackets with subscript $\theta$ denote volume-weighted averages in the $\theta$-coordinate, and angle brackets with $t$ subscript indicate a time-weighted average. 
%The convective region extends, in terms of radius, from  $r_{\rm core}$ to $r_{\rm surf}$, the top of the convective zone.
%The term being time averaged, is an average of $\langle\tau_{conv}\rangle_\theta$ over the convective region from $r_{\rm rad.}$, the top of the radiative zone, to $r_{\rm surf.}$, the top of the convective zone weighted by the radial shell spacing $dr$. 
%The values of $\tconv$ for the various simulations are given in Table~\ref{table:sim-sum} and show that time averages over  $10^{7}$~s represent $\sim$ 17 to 26 convective turn over time scales depending on the simulation.
%Knowing the convective turn over time scales of the various simulations the $10^{7}$~s time averages of $\left|v\right |_{\rm rms}$ in Figure~\ref{fig:vrms-1e-4} have been averaged over 17 to 26 convective turn over time scales depending on the simulation, with fewer convective turn overs for the higher $\eacc$ simulations and more convective turn overs for the lower $\eacc$ simulations.

%The top panel of Figure~\ref{fig:vrms-1e-4} shows the time averaged angular rms of the velocity magnitude, while the middle and bottom panels show the time averaged angular rms of the $r$ and $\theta$ velocity components. 
%Figure~\ref{fig:vrms-1e-4} shows the time and angular averaged of the rms velocity magnitude. 
Note in figure~\ref{fig:vrms-1e-4} that the large values of the rms velocities near the surface are due to  the tangential component of the velocity of the flows that develop close to the outer boundary.  In the following, we refer to ``cold", cases with entropy of the accreted material lower than the bulk entropy in the convective envelope (e1 - e4), ``moderately hot" if the accreted entropy is slightly higher (e5-e7)
and ``hot" cases if the accreted entropy is much higher (e8-e9).
 Cold cases to moderately hot accretion models tend to show a slight increase in convective velocities relative to the non-accreting model (black curve). Hot accretion cases (e8, e9) show a decrease in convective velocities in the bulk of the convective region. This decrease is also evident from the larger values of the convective turnover timescales given in Table~\ref{table:sim-sum}.
 The radial profile of the rms of the {\it radial} velocity component shows the same trend  as a function of the accreted entropy.
 % as displayed by the rms of the velocity magnitude  in Fig. ~\ref{fig:vrms-1e-4}.
%It is quite obvious from this plot and the values of the convective turn over time scale given in Table~\ref{table:sim-sum} that the e9 case shows a decrease in convective velocity in the bulk of the convective region relative to the non-accreting model. Cooler accretion simulations (e1 -- e7) tend to show a slight increase in convective velocities relative to the non-accreting model. Case e8, like case e9, also shows a decrease in convective velocities though not as large a decrease as the e9 case. This is also evident in the convective turn over time scale given in Table~\ref{table:sim-sum} for the e8 case. %While the e9 case shows a decreased convective velocity in the bulk of the convective region, it shows an increase of velocity in the outer 10\% of the model by radius. By comparing the radial and angular velocity components one can see that the majority of the increased velocity magnitude comes from the angular component and there is actually a decrease in the radial component relative to the non-accreting model. Interestingly case e8, which shows a decrease in convective velocity in the bulk of the convective region similar to case e9, does not show the increase in the angular velocity component in the outer 10\% of the model and is more similar to the lower energy accretion calculations in the shape of the $\left|v\right|_{\rm rms}$ near the surface.

In addition to the convective velocities, an analysis of the enthalpy luminosity for the different simulations gives insight into the role of convection.  
The mean radial enthalpy luminosity in the convective zone is related to the energy transported by convection and is equivalent to the convective luminosity in the framework of the mixing-length theory. It is computed as 
\begin{equation}
L_{\rm h} (r) = \langle \, \langle \sr \ur \rho (e + { P \over \rho}) \rangle_\theta - \langle \sr \ur \rho \rangle_\theta \langle  (e + { P \over \rho})  \rangle_\theta \, \rangle_{t},
\end{equation}
where $P$ is the pressure and $\sr$ is the surface area of a cell at radius $r$ \citep[see][]{Viallet-2011}. The second term on the r.h.s. of the equation accounts for non-zero mean vertical mass fluxes  \citep{Freytag-1996}, which can be important in the case of accretion.
%Note that $F_{\rm h}$ has the dimension of a flux multiplied by a surface in order to take the geometrical effects introduced by the spherical geometry into account \citep{Viallet-2011}. It thus corresponds to a luminosity and will be referred to the enthalpy luminosity in the following.
Figure~\ref{fig:lenthalpy} shows radial profiles of $L_{\rm h}$, time-averaged over $2 \times 10^7$ s, for the non-accreting (mdot0),  cold (e1), and hot (e9) accreting cases.  The behaviour of the enthalpy luminosity is similar to that found for the rms velocities. Both quantities show a significant decrease in the bulk of the convective zone for the hot accretion case (e9) compared to the non-accreting (mdot0) and cold accreting (e1) cases. Heating the surface layers of a convective star, as observed in Figure~\ref{fig:temp-struct} for the hot accretion cases, leads to a reduction of the rate of energy transport by convection 
%and thus a decrease in the convective velocity and flux for the hot accretion case e9, 
compared to the cold and non-accreting cases. Surface cooling and steep temperature gradients near the surface drive convection in a stellar envelope. Adding heat to the surface  locally changes (or even inverts) the temperature gradient, and therefore can inhibit convection.  
Despite the suppression of convection, calculation of the mean radiative luminosity $L_{\rm rad}$  as a function of radius\footnote{See \cite{Viallet-2011} for the calculation of  the mean radiative luminosity.} shows that it remains by several orders of magnitudes smaller than $L_{\rm h}$ in the bulk of the convective zone for all simulations. This stems from the fact that the thermal profile and thus the radiative flux require a thermal relaxation timescale to adjust and eventually to compensate for the reduced efficiency of convection.  
An estimation of the thermal relaxation timescale of the upper layers (see \S \ref{1D}) indicates  that only layers with $r/R > 0.99$ have enough time to adjust during the total time of the simulations, i.e.  $ 3 \times \,10^7$\,s. Our interpretation and the consequences of the results are further analysed in the next sections.

%Our interpretation of the results is that the significant heating of the surface layers yields to their expansion, contributing negatively to the total luminosity and counterbalancing the positive contribution which arises from the deeper contracting regions. 
%TODO: say a bit more about the vorticity plot
%TODO: talk about balance between e_acc and L

\section{Distribution of Accreted Material}
\label{distribution}

\subsection{Numerical mass distribution }

\begin{figure}[h]%10
\centering
\includegraphics[width=\singlecolumnfigurewidth]{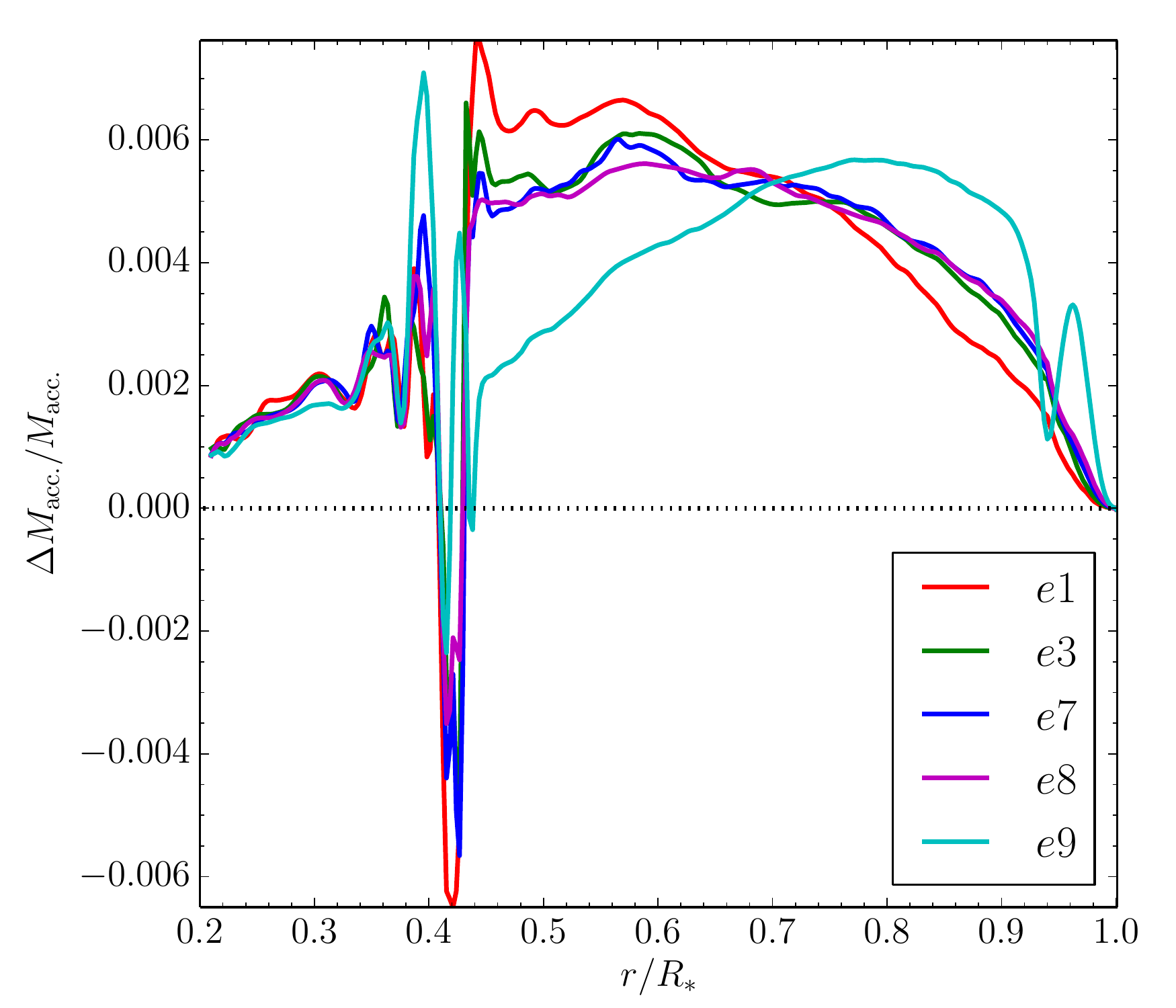}
\caption{Change in mass distribution in each radial shell of an accreting model relative to the non-accreting model for different values of $\eacc$ at time $t=3\times10^7$~s. The reference model used to calculate $\Delta M_{\rm acc}$ is the non-accreting model at same age (see text, \S \ref{distribution}).}
\label{fig:mass-dist}
\end{figure}

%\begin{figure}[h]%11
%\centering
%\includegraphics[width=\singlecolumnfigurewidth]{./figures/f11.pdf}
%\caption{***DELETE FIGURE***Variation of the mass distribution shown in Figure~\ref{fig:mass-dist} with time for the e1 (top panel) and e9 (bottom panel) models.}
%\label{fig:mass-dist-time}
%\end{figure}

%One of the main goals of this work is to start to understand how the accreted material is distributed through out the stellar model. 
An additional interpretation of the results is provided by analysing how
the accreted material is distributed throughout the stellar interior. 
We first approach the problem by comparing at a given time the mass in each radial shell to that in each corresponding shell in an otherwise identical non-accreting model. The result of this comparison is shown in Figure~\ref{fig:mass-dist} where the change in the mass of each shell $\Delta M_{\rm acc}$, derived from the change in density, is expressed as a fraction of the total accreted mass $M_{\rm acc}=\dot{M}\Delta t$.  The quantity $\Delta M_{\rm acc}$ is calculated  from the density and volume of the computational cells in a given radial shell (i.e. averaged over $\theta$) and for a model at a given time (not time-averaged). 
Note that the sum of $\Delta M_{\rm acc}/M_{\rm acc}$ over all shells in the stellar model is equal to 1.
%Plotted is the ratio of the change in mass due to accretion, $\Delta M_{\rm acc}$, calculated  from the density and volume of the computational cells in a given radial shell (i.e averaged over $\theta$) to the total mass accreted, $M_{\rm acc}=\dot{M}\Delta t$. 
%This measure of the change in density relative to the non-accreting model accounts for the addition of newly accreted material as well as compression of original stellar material as a result of the accretion process. 
The change in density of a cell comes from i) compression due to the addition of material in layers above, and ii) from the addition of the newly accreted material.
Thus, the quantity $\Delta M_{\rm acc}/M_{\rm acc}$ is a measure of the net change in stellar density as a result of the accretion process 
and not just a measure of the mass added in each shell. 
%We will attempt to address this difference later in this section.  
Figure~\ref{fig:mass-dist} shows the similarity of results with different accretion entropy in the range $ -30\% \leq \delta s_{\rm acc} \leq 30\%$ (cases e1 to e7).  Higher accretion entropy, however, favours a density increase closer to the surface of the star.  This trend is clearest in the hottest case, e9. We stress that the curves plotted in Figure~\ref{fig:mass-dist} and linking quantities that are defined in each shell are shown for the sake of illustrating the differences in the mass distribution between different accretion energy cases. While the details will depend on the resolution, the overall trends will not.
%Clearly these quantities  and the curves  will highly depend on the resolution.

%Considering again the mass distribution shown in Figures~\ref{fig:mass-dist}, if one performs an integral of $\Delta M_{\rm acc}/M_{\rm acc}$ over the star it sums to 1.0, indicating that the total mass accreted is correctly accounted for in this measure. 
%The increase in density of a cell can come from i) compression due to the addition of material in layers above, and ii) from the addition of the newly accreted material. %If one calculates the fraction of accreted mass in a radial shell purely from the density of the stellar model these two sources of increased density are indistinguishable from each other and only the sum is seen. This is what Figure~\ref{fig:mass-dist}~and~\ref{fig:mass-dist-time} show. 
%With this picture of what $\Delta M_{\rm acc}/M_{\rm acc}$ is measuring the non-zero fraction in the radiative zone is more intuitive, rather than suggesting that some accreted material is mixing into the radiative region it is likely the result of additional compression resulting from extra mass being added to the overlying layers. However, using only $\Delta M_{\rm acc}/M_{\rm acc}$ it is not entirely clear wether any accreted material is indeed entering the radiative region or not.

To better understand the difference between density variation due to compression of material and that due to mass addition, we use another method, namely tracer particles. The non-interacting particles trace out the path an infinitesimally small fluid element would follow with a given initial position subject to the velocity field calculated by MUSIC. We use a simple predictor-corrector method to integrate the particle positions. 
%In our case we have all the primitive hydrodynamic quantities at two times $n$ and $n+1$. 
 In order to determine each tracer particle's motion between two timesteps, say $n$ and $n+1$, 
we first estimate the particle position at time $n+1/2$  based only on the velocity field at time $n$. We then use this half-way position to find a more representative velocity for the whole time interval of the particle's motion between the two timesteps. 
%We use cubic spline interpolation for the velocities in space and linear interpolation of the velocities in time to determine the value of velocity field at the half way location of the tracer particle at the centred time between the two time states. Cubic spline is a reasonably accurate interpolation method, and the use of any higher order interpolation is s not justified. A linear interpolation in time in used since only two time states are saved in MUSIC to limit the memory usage of the code.

\begin{figure}[h]%13
\centering
\includegraphics[width=\singlecolumnfigurewidth]{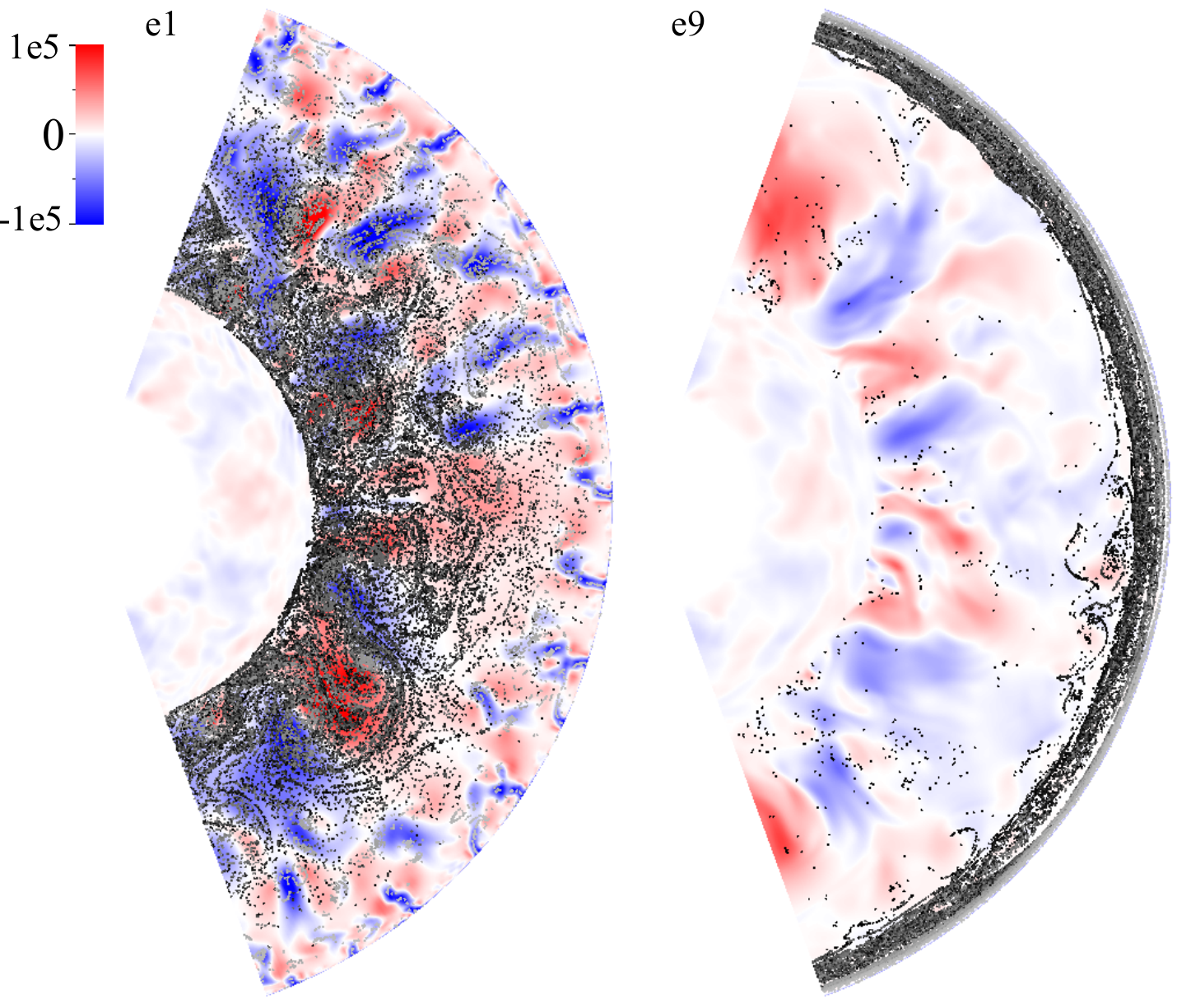}
\caption{Distribution of tracer particles accreted during $1.5\times 10^7$~s for the low energy (e1) and high energy accretion (e9) cases respectively. Colour scale is the radial velocity in units of cm s$^{-1}$ and grey scale points show the position of the tracer particles. Velocities are instantaneous, i.e. not time-averaged. Blue indicates down flows, while red indicates up flows. Lighter coloured points correspond to more recently added tracer particles. Both snapshots are at a time  $t=1.5\times 10^7$~s}
\label{fig:mass-v_r-part}
\end{figure}

 To study the distribution of the accreted material we add the particles over time at the surface of the star at a fixed rate
and over the accreting area $\Aacc$. %Given the fixed rate at which these particles are added and the fixed mass accretion rate 
%One can associate a mass to each particle to represent the amount of matter accreted.  
%However, one must keep in mind that while the tracer particles may give a representative measure of where the accreted material could mix, they are not directly attached to the accreted material. Only through the sharing of a common velocity field are the two processes linked. A potentially better, and also more computationally demanding method, would be to compute the advection of a passive scalar associated with accretion. However, doing such a study is beyond the scope of the current paper. By 
Examining the radial distribution of particles gives some insight into how the newly accreted material is distributed.  Figure~\ref{fig:mass-v_r-part} shows the result of adding 75,000 tracer particles,  at 200\,s intervals and  randomly distributed in the $\theta$-direction within the accreting region, 
from the start of accretion until $1.5\times 10^{7}$~s.  
The grey scale applied to the particles indicates when they were added to the simulation, with darker particles being added earlier than lighter particles. The colour scale indicates the radial velocities. 
Motions characteristic of convection \citep[see e.g.][]{Stein-1998} can be seen with down flows, indicated by blue, and up flows, indicated by red. 
%Bluer regions show down flows, while redder regions show up flows. 
The left panel of the figure shows the distribution of particles in the low energy (e1) case, while the right panel shows the distribution of particles in the high energy (e9) case. In the low energy case the particles are distributed throughout the convective zone. In the high energy case they are concentrated in the outer 10\% of the stellar radius.  The few particles that  penetrate deeper into the convective region in the latter case were added early in the accretion process and transported before the structure, subjected to the  new accretion boundary condition, reached a steady state. 
% (as in Figures~\ref{fig:density-struct}~and~\ref{fig:temp-struct}).

\begin{figure}%14
\centering
\includegraphics[width=8cm]{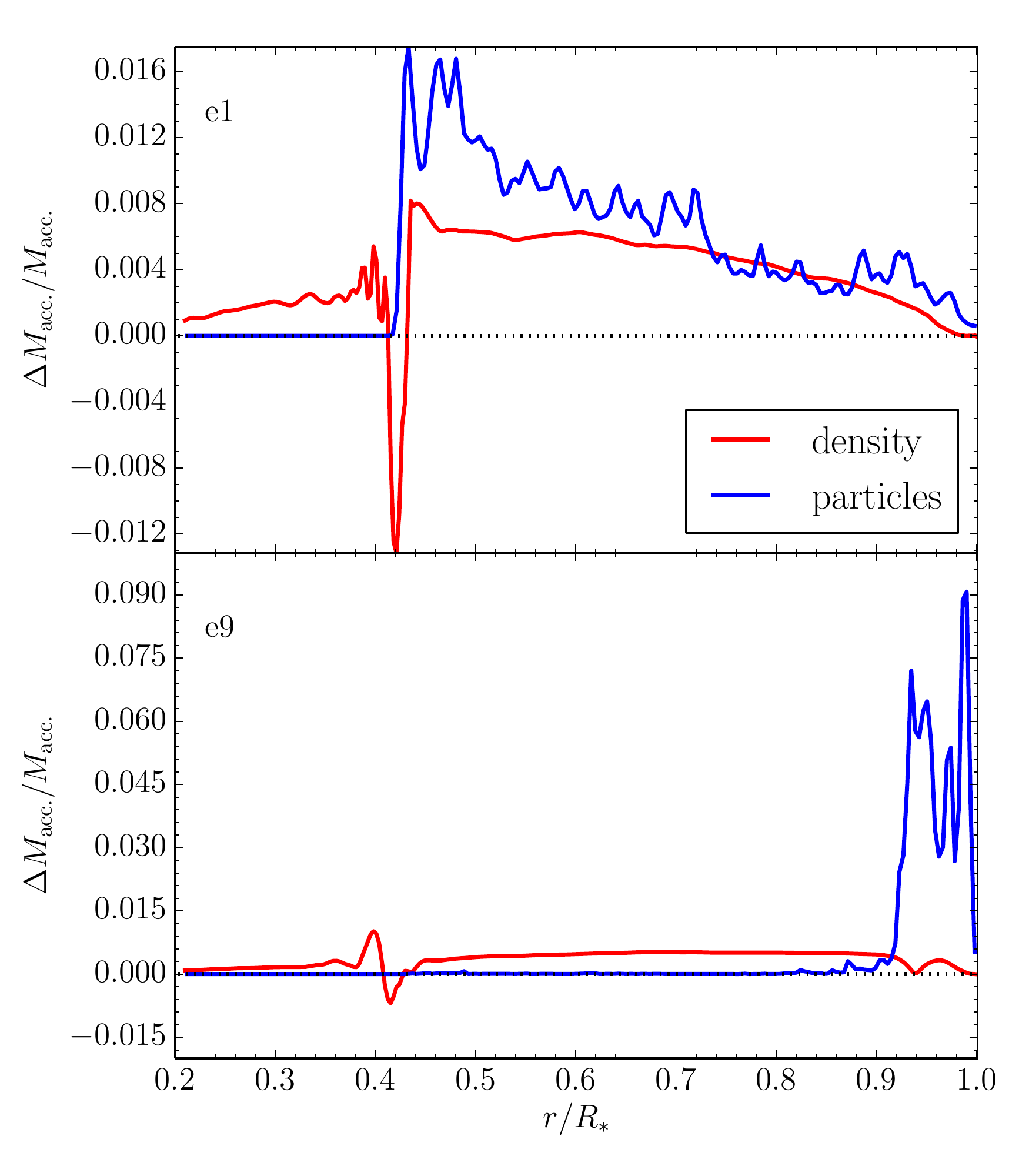}
\caption{Comparison of mass distributions between that derived from model density and that from  particle distributions for the e1 (top panel) and e9 (bottom panel) models. The mass distributions are derived from the models displayed in  Fig. \ref{fig:mass-v_r-part} at time $t=1.5\times 10^{7}$~s.}
\label{fig:mass-dist-part}
\end{figure}

One can associate a mass to each particle to represent the amount of matter accreted, given the fixed rate at which these particles are added and the fixed mass accretion rate. 
 We can thus derive a distribution of accreted mass from the particles. This mass distribution 
 is shown in Figure~\ref{fig:mass-dist-part} and is compared to the accreted mass distribution based on the change in density. Note that  the mass distribution for the e1 case plotted in Figure~\ref{fig:mass-dist} (red curve) slightly differs from the one plotted in Figure~\ref{fig:mass-dist-part} since they are not evaluated at the same time. The particle derived accreted mass distributions do not take into account the compression which occurs due to accretion. They should thus not necessarily match that derived from the density change. 
%The particles represent only one component of the mass in the cell, the accreted mass, and says nothing about how the original stellar material may have been compressed. 
%It also says nothing about how the original stellar material may have been redistributed before it has mixed with the accreted material, as no tracer particles were added to the regions where only original stellar material existed. 

In the e1 case the two methods based on density change and on particles respectively, show a similar distribution, however, the particles reveal that there is no mixing of accreted material in the radiative region. 
%This is not surprising as those low resolution simulations do not show obvious penetration of convection into the radiative region. 
%The two different methods of measuring the distribution of accreted material are really measuring different things. The particles are telling use about how the accreted material will likely mix, while the change in density of the computational cells is telling us about the net effect accretion has on the density of the star.  In the e9 case this is even more evident. 
In the e9 case, the particle derived accretion mass distribution shows that the newly accreted material remains near the surface and does not mix into the deep convective region.  This is consistent with having high entropy material above lower entropy material. The two mass distributions derived from the particles and from the density change are thus different, as shown in the bottom panel of Figure~\ref{fig:mass-dist-part}. 
%while in the e1 simulation, except for the radiative region, the two distributions follow each other more closely. 
%This also comforts us with the interpretation that the density change in the radiative region is a result of accretion through compression of the original stellar material and not of mixing of the newly accreted material into the bulk of the radiative region.

\subsection{Comparison to the analytical model of \cite{Siess-1996}}

The only work we know that attempts to model the mass distribution resulting from accretion in stellar interiors  is the work of  \cite{Siess-1996}. These authors construct a complex analytical model to specify the accreted mass distribution in the framework of one dimensional stellar evolution models. It is interesting to compare predictions from their model to the results of our numerical experiments. \cite{Siess-1996} define a radially dependent penetration function, $f=\Delta M_{\rm acc}/\Delta M_*$, where $\Delta M_{\rm acc}$ and $\Delta M_*$ are  the accreted and initial mass in a given radial shell, respectively. The computation of
$f$ results from the integration of a differential equation \citep[see Eq. (23) in][]{Siess-1996} that
%using the 4$^{\rm th}$ order Rung-Kutta method integrating inward from the surface of the stellar model. The differential equation for the penetration function 
%includes terms to account for composition gradients, which we do not have as our models have a uniform composition, so these terms do not contribute. The differential equation also 
depends on quantities derived from the stellar model such as the pressure scale height, adiabatic gradient, and stellar radius and mass. In addition, the model contains several free parameters: $\xi$ is the fraction of the Keplerian specific angular momentum of the accreted material, $Ri^-$ is the Richardson number, which must take negative values in convective regions, and $\epsilon^a$ is related to how quickly the accreted matter is thermalised \citep[see][for details about these parameters]{Siess-1996}.

\begin{figure}[h]%12
\centering
\includegraphics[width=8cm]{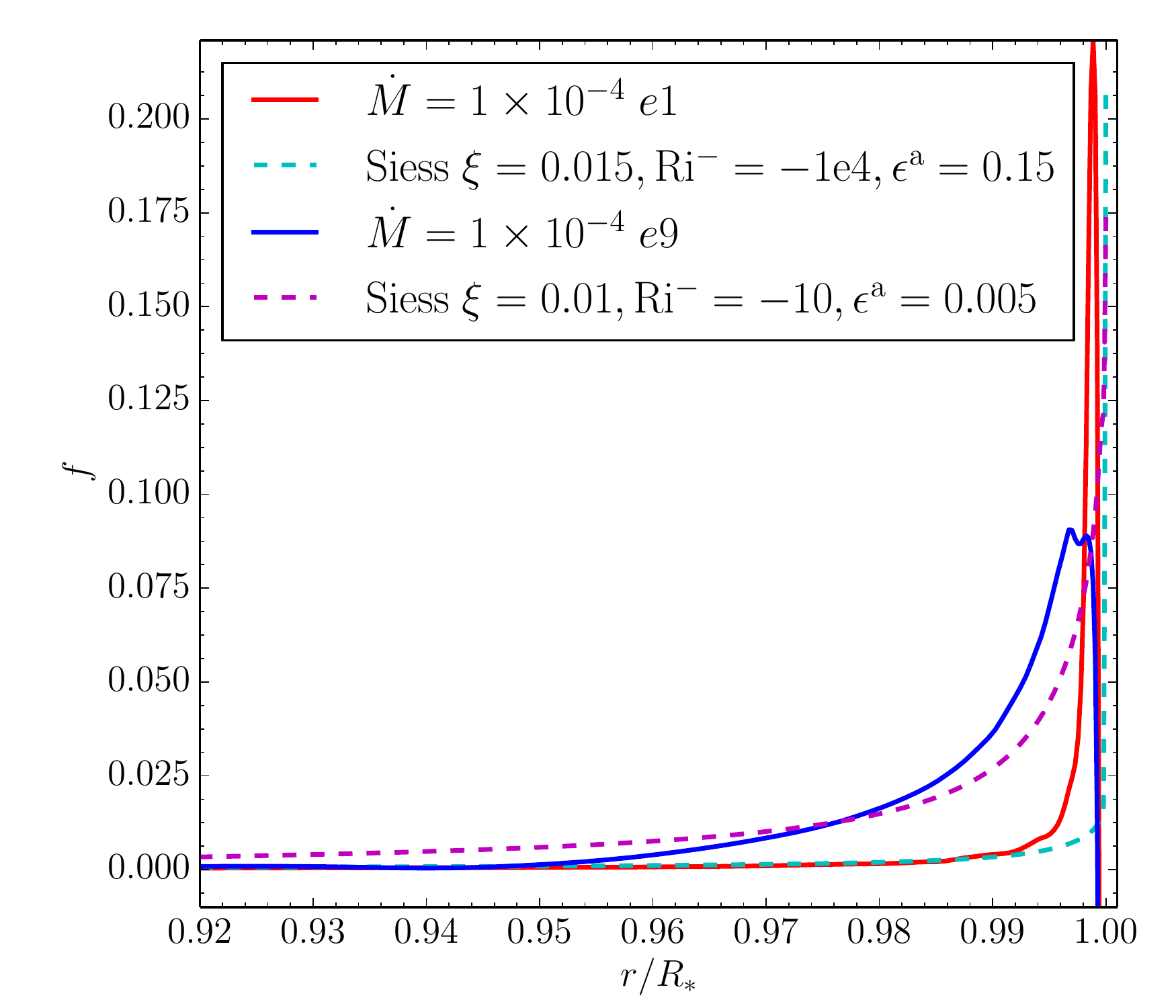}
\caption{Comparison between penetration functions derived from our 2D simulations and that from the analytic model of \cite{Siess-1996}. The figure shows a comparison for the e1 simulation (solid red curve) with our best fit analytic function (dashed cyan curve) and a comparison for the e9 simulation (solid blue curve) with our best fit analytic function (dashed maroon curve). %The bottom panel shows how better fitting the curve shape in the interior compromises the fit near the surface for a lower mass accretion rate calculation of $10^{-6}$~M$_\odot$~yr$^{-1}$.
}
\label{fig:pen-func}
\end{figure}

We can compute the equivalent of the penetration function $f$ directly from our 2D simulations of accretion,
because they provide the quantities $\Delta M_{\rm acc}$ and $\Delta M_*$. 
This numerical penetration function can be
compared to the semi-analytical values of $f$ obtained from solving the differential equation of \cite{Siess-1996}.
The quantity $\Delta M_{\rm acc}$ for the 2D simulations is derived from the density change  and not from the particles
since the latter, as already mentioned, only partially accounts for the effects of mass addition, omitting the effect of compression due to the addition of material. The former derivation thus provides a more realistic description of the net effect of mass accretion, which is what should count in a 1D stellar evolution calculation using a penetration function. 
% using a 4$^{\rm th}$ order Rung-Kutta method integrating inward from the surface of the stellar model and using angular-averaged quantities as a function of the stellar radius. 
The results are shown in Figure~\ref{fig:pen-func} for the e1 and e9 accretion simulations. The dashed cyan ($\xi=0.015$, Ri$^{-}=-1\times 10^4$, $\epsilon^a=0.15$) and maroon ($\xi=0.01$, Ri$^{-}=-10$, $\epsilon^a=0.005$) curves are our best attempts to fit the numerical results for the e1 and e9 cases respectively. The analytic penetration functions are poor fits to those derived from the 2D simulations. Despite many attempts, we find it difficult to match both the peak of the curve at the surface and the shape of the curve below the surface. Interestingly though,  the best fit for the e9 case requires a smaller absolute value for Ri$^{-}$ than that needed for the e1 case. Small absolute values of this parameter correspond to a regime where convection is not very efficient, while large absolute values indicate efficient convection \citep{Siess-1996}, in line with our findings in \S \ref{structure}. 

%In fact, because the penetration function is divided by $\Delta M_*$ much of the departure away from zero deeper in the model is not noticeable and so matching the penetration functions this way is only sensitive to the surface. 
Experimenting further, we note that we are able to better fit the shape of the penetration function for a lower accretion rate. We performed a 2D simulation with a mass accretion rate $\dot{M}=10^{-6}$~M$_\odot$~yr$^{-1}$ and low accreted entropy as in the e1 case. The corresponding analytical penetration function can fit the numerical results with a set of free parameters ($\xi=0.01$, Ri$^{-}=-2.2 \times 10^3$, $\epsilon^a=0.15$).
The values of $f$ very close to the surface remain, however, significantly higher than computed from the 2D simulation. We do not wish to push further such comparisons, which certainly remain limited as our numerical simulations do not account for deposition and inward transport of angular momentum. But the main lesson  from such comparisons is that the \cite{Siess-1996} formalism, in addition of being complicated for implementation in 1D stellar evolution code,  requires fine tuning of the free parameters depending on the accretion details, and may break for extreme cases of high accretion  and energy deposition rates. 
%The bottom panel of Figure~\ref{fig:pen-func} shows a fit to a lower accretion rate simulation with a mass accretion rate of $\dot{M}=10^{-6}$~M$_\odot$~yr$^{-1}$. In this case we were able to fit the shape of the penetration function better, however, the values very close to the surface are significantly higher than in the 2D simulation.

\section{Impact on one dimensional structure and evolution}
\label{1D}

The results from the 2D simulations presented in the previous sections can be used as guidance for the treatment of accretion in 1D stellar evolution calculations. They show that the redistribution of the accreted material within the convective envelope remains efficient for cold and even moderately hot accretion (e1 to e7 cases). These cases correspond to values of $\alpha$ (see Eq. (1)) ranging from $\sim$ 0.01 (e1) to $\sim$ 0.06 (e7).
A  different behaviour in terms of mass redistribution,  rms velocities in the convective envelope and enthalpy flux is observed for the hottest cases investigated (e8 and e9), corresponding to values of $\alpha$ of $\sim$ 0.1 for e8 and $\sim$ 0.2 for e9. Note that,  for a given initial stellar model, the exact value of $\eacc$ at which this change in behaviour is observed will likely  depend on the accretion rate, the accreting surface area, the inflow velocity (see \S \ref{accretion}), as well as the  boundary conditions and spatial resolution. Thus, the quantitative value of  $\eacc$ (or $\alpha$) at which this change in behaviour occurs is not very significant, only the qualitative effects highlighted by the 2D simulations is of importance.
The visualisation, thanks to the tracer particles,   of a hot layer produced at the surface of the accreting object due to the accumulation of high entropy material (see Figure\ref{fig:mass-v_r-part}) raises several questions. 
First, what is the long-term effect of such a hot buffer zone on the structure and evolution of the accreting object? Second, since obviously the hot accreted material is prevented from sinking into the deeper layers, is the assumption
of redistribution of accreted energy into the interior \citep{Siess-1997, Hartmann-1997, Baraffe-2012}
able to mimic the effect of this hot layer at the surface? 

We address these questions with our 1D Lyon stellar evolution code, assuming different treatments of accretion. The accreted mass is assumed to be redistributed over the entire structure, whether accretion is cold or hot \citep[see Eq. (3) of][]{Baraffe-2010}\footnote{Note that the Lyon stellar evolution code is based on the Lagrangian coordinate $m_{\rm r}$, i.e. the mass enclosed in a sphere of radius $r$, and needs to account for 
the change in entropy due to mass variation during the accretion process  as done in e.g. \citet{Hayashi-1965, Baraffe-2010}.}.
Starting from the 1D young Sun structure used for the initial model in the 2D simulations, we run 1D simulations with a constant accretion rate of $\dot M = 10^{-4} \msolyr$ for approximately 100\,yr. This corresponds to the typical duration of high accretion bursts 
adopted in the framework of episodic accretion scenarios \citep{Hosokawa-2011, Baraffe-2012}. 
On one hand, we adopt a similar treatment to that described in  \cite{Baraffe-2009, Baraffe-2010} assuming cold accretion, equivalent to $\alpha=0$, and a hot accretion case assuming uniform redistribution within the stellar interior of the accretion energy, $\Ladd$ (see Eq.~(1)), with $\alpha=0.2$ to correspond to the e9 case. 
The $\alpha=0.2$ case is relevant for the scenario developed by \cite{Baraffe-2012} to explain some of the observed properties of FU Ori. Note that such a treatment assumes that the accreting star can freely radiate its energy over most of its photosphere \citep[see also][]{Hartmann-1997}. 
%Note also that such treatment accounts for the change of entropy due to mass variation during the accretion process \citep{Hayashi-1965, Baraffe-2010}.
On the other hand, we adopt the treatment inspired by the results of the 2D simulation for the hottest case and impose the boundary condition that the surface luminosity $L_{\rm surf}$ equals the accretion energy rate $\Ladd$.  This treatment is similar to the one adopted by \cite{Mercer-1984}. In this case, only the outer boundary condition is modified and no accretion energy is redistributed within the stellar interior.
This forces the value of the effective temperature (the surface temperature $T_{\rm surf}$) to  tend toward the value $(\Ladd/(4 \pi \sigma R^2))^{1 \over 4}$.  
It is also equivalent to the boundary condition used at the top of the domain of the 2D simulations, where it is assumed that the surface radiates a flux $F = \sigma T^4$  with $T$ the temperature of the cells at the top of the domain (see \S~\ref{initial}). 
 Adopting  $\alpha=0.2$ yields $\Ladd \sim 100\,L_\odot$, a value comparable to the surface luminosity of the e9 case displayed in Table~\ref{table:sim-sum}.

\begin{figure}[h]%6
\centering
\vspace{-1cm}
\includegraphics[width=8cm]{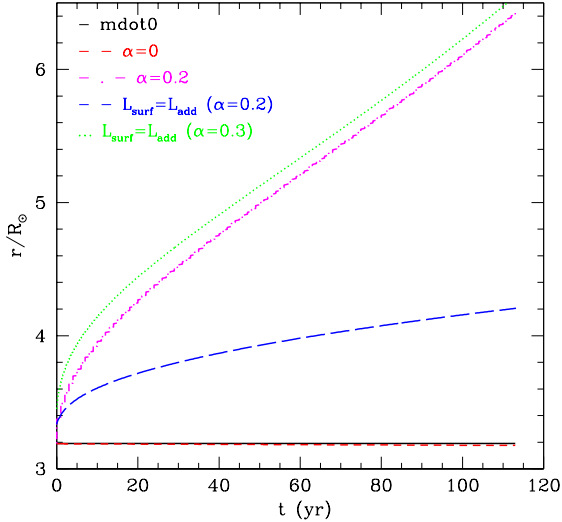}
\vspace{-2cm}
  \caption{Evolution of the radius of non-accreting and accreting young Sun models calculated with the 1D stellar evolution code.
 The non-accreting model  is shown as the black line. Coloured curves indicate accreting models with a rate $\dot M = 10^{-4}\,\msolyr$ assuming different treatments of accretion: ``cold" accretion (red dash); ``hot" accretion assuming uniform redistribution within the stellar interior of the accretion energy rate $L_{\rm add}$ with $\alpha=0.2$ (magenta dash-dot ); ``hot" accretion assuming an accretion boundary condition $L_{\rm surf}=L_{\rm add}$ with $\alpha=0.2$ (blue dash) and $\alpha=0.3$ (green dot).}
\label{fig:tr1D}
\end{figure}

The results displayed in Figure~\ref{fig:tr1D} show the evolution of the radius in the 1D models with two different treatments of hot accretion: uniform energy redistribution and outer accretion boundary condition. In both cases the two different treatments of hot accretion yield  a rapid expansion of the accreting object, with the expansion first proceeding  in the outer surface layers close to the photosphere, given their much shorter relaxation timescale compared to the deeper layers. Figure~\ref{fig:tr1D}
shows that for the same rate of energy added to the object through accretion, $\Ladd$, a model assuming uniform redistribution of the energy within the interior (dash-dot magenta curve) overestimates the growth in the radius compared to a model assuming a more realistic accretion boundary condition where hot accreted material cannot sink and produces a hot surface layer (long dash blue curve). In order to obtain a similar evolution with an accretion boundary condition as obtained with the assumption of energy redistribution within the interior,  one needs to increase by a factor 1.5 the amount of accretion energy transferred to the star, corresponding to a value of $\alpha=0.3$ (green dotted curve in Fig. ~\ref{fig:tr1D}).

Imposing an accretion boundary  yields rapid heating of the photosphere and of  the layers below. This heating results in the expansion of the outer layers, which starts after  $\sim$ 1\,yr of evolution.  The heating and thus the modification of the thermal profile propagates from the surface to deeper layers on a thermal relaxation timescale $\tth$, which can be estimated by
\begin{equation}
\tth \approx {< c_{\rm P} T>_{\Delta m}  \Delta m \over L},
\label{tth}
\end{equation}
where $\Delta m$ is the mass of the layers considered below the surface,  $c_{\rm P}$ the specific heat at constant pressure and  the brackets $<>_{\Delta m}$ denote the  average over these layers. With the boundary condition that  $L = \Ladd$ with $\alpha=0.2$,  modification of the thermal profile in the corresponding 1D simulation is observed for the very top layers with a fractional mass $m_{\rm r} / M \simgt 0.9998$ after about 1\,yr of evolution. This is consistent with the estimate of the thermal timescale of these layers $\tth \sim$ 1\,yr derived from Eq. \ref{tth} and with the change in the thermal profile  illustrated in Figure~\ref{fig:temp-struct} for the 2D simulations. 
After around 100\,yr of evolution, the 1D model with the accreting boundary condition shows a modification of its thermal profile for layers down to a depth $m_{\rm r} / M \sim 0.985$, consistent with their thermal timescale $\tth \sim 150$\,yr. The expansion of models with a hot accretion condition displayed in  Figure~\ref{fig:tr1D} stems from the expansion of these upper layers characterised by a fast adjustment timescale.
Note that the thermal timescale of the whole star, given by the Kelvin-Helmotz timescale $\tau_{\rm KH} = G M^2/(R L)$, is around $4$\,Myr for the initial young Sun (non-accreting) model and about $0.1$\,Myr for the accreting model with the accretion boundary condition and $\alpha=0.2$ (it is shorter in the latter case because of the higher imposed surface luminosity).

\begin{figure}[h]%6
\centering
\vspace{-1cm}
\includegraphics[width=9cm]{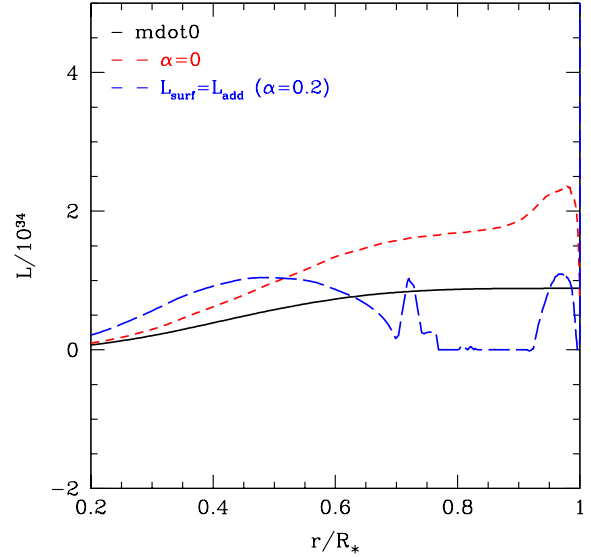}
\vspace{-2cm}
  \caption{Radial profile of the  luminosity (in erg s$^{-1}$) in the young Sun model after 1\,yr of evolution, based on 1D stellar evolution calculations. The black solid line represented the non-accreting model. Coloured curves indicate accreting models with a rate $\dot M = 10^{-4} \msolyr$ assuming  ``cold" accretion (red dash-dot) and ``hot" accretion with an accretion boundary condition $L_{\rm surf}=L_{\rm add}$ with $\alpha=0.2$ (blue dash) .}
\label{fig:rl1D}
\end{figure}

To further compare with the 2D results, we analyse the total luminosity profile as a function of radius for 1D models in Figure~\ref{fig:rl1D} after 1\,yr of simulation starting from the same initial model. This  is similar to the total time spanned by the 2D simulations with accretion ($ 3 \times 10^7$ s). The non-accreting model (black curve) shows a profile characteristic of a pre-Main Sequence star contracting as a whole and with no nuclear energy source, with the luminosity $L$ powered by local gravitational energy due to contraction $-P {dV \over dt}$ and by change in internal energy  $-{de \over dt}$:
\begin{equation}
L= \int_M - \left ( P {dV \over dt} + {de \over dt} \right) \,dm.
\end{equation}

 The total luminosity in the convective envelope is representative of the convective luminosity, because energy transport is dominated by convection and the contribution from radiation is negligible in the non-accreting model. 
The cold accretion case shows a similar behaviour, but  in addition to the contraction of the whole object, there is also a contribution to the gravitational energy due to the mixing of cold entropy material in the interior, as visualised in the 2D model for the e1 case in Figure~\ref{fig:mass-v_r-part}. As shown in Figure~\ref{fig:rl1D}, this yields  a higher interior luminosity compared to the non-accreting case. An equivalent result is found in the 2D  cold and moderately hot accretion cases. This behaviour observed in the 1D models explains the higher enthalpy luminosity in the 2D e1 model relative to the 2D non-accreting model  (compare the e1 case to the non-accreting case  in Figure~\ref{fig:lenthalpy}).
For the hot accretion case based on the accretion boundary condition (blue dashed curve in Fig. ~\ref{fig:rl1D}), the luminosity significantly decreases within the bulk of the convective zone, as observed  for the 2D e9 case in   Figure~\ref{fig:lenthalpy}. 
Inspection of the different terms contributing to the luminosity in the 1D model indicates a slight expansion of the outer envelope, i.e. $- P {dV \over dt} < 0$, contributing negatively to the luminosity and counterbalancing the positive contribution from the deeper contracting regions.  Such contributions can already be seen after 1 yr of evolution, even if the change in total radius remains small.   As the 1D model with the accretion boundary condition further evolves and expands, examination of its interior structure shows that the top boundary of the convective zone recedes from the surface, as also noted by \citet{Prialnik-1985}. Additionally, in the convective zone, the energy transport by convection becomes less efficient and the contribution of radiative transport increases. 

The parallel between the 1D and 2D results is interesting. First, it indicates that the 1D treatment of cold accretion assumed in previous studies is roughly consistent with the 2D results, with efficient redistribution of the accreted mass within the interior and similar effect of accretion on the luminosity in the convective envelope. Second, for hot accretion, the assumption  of redistribution of the accreted energy within the interior seems to overestimate the effect on the structure, namely the expansion of the accreting object, for a given total amount of accretion energy gained. An accretion  boundary condition assuming $L_{\rm surf} = \Ladd$ seems more realistic and more in line with the 2D results. However, despite the quite unphysical picture represented by the assumption of accreted energy redistribution within the interior in the case of hot accretion, it can mimic the effect described by the 2D simulations, assuming a smaller amount of the accretion energy  is transported deep into the interior.

\section{Discussion and conclusion}
\label{discussion}

Our main results can be summarised as follows.  The usual assumption made in 1D stellar evolution codes of instantaneous and homogeneous redistribution of accreted material within the interior of the 
accreting object is supported by the multi-D simulations in the case of low to moderately high entropy accreted material. 
In this context, ``low" or ``high" entropy refers to the value of the entropy of the accreted material compared to the entropy of the adiabatic convective envelope.
For high entropy material, the assumption of energy redistribution within the interior is physically wrong and opposite to results depicted by the multi-D simulations, namely there is an accumulation of hot material at the surface that does not sink and instead produces a hot surface layer. The presence of this hot surface layer tends to suppress convection in the envelope.
 An accretion boundary condition such as $L_{\rm surf} = \Ladd$, in the case of ``hot" accretion, provides a more physical treatment. For  a given amount of accreted energy transferred to the accreting object, a treatment assuming accretion energy redistribution throughout the stellar interior significantly overestimates the effects on the stellar structure, and in particular on the resulting expansion. 
 
 Given the lack of knowledge regarding the amount of accretion energy transferred to the accreting object, a systematic analysis is required to explore the impact of these results. Work is in progress 
 to understand how the use of the accretion boundary condition inspired by this work will affect the conclusions of previous studies based on the former treatment for hot accretion \citep{Baraffe-2009, Hosokawa-2011, Baraffe-2012} and the interpretation of observations relying on hot accretion scenario \citep[e.g.][]{Hartmann-2011}.   
 %Interestingly however, a 1D treatment assuming energy redistribution can mimic the effect of such an accretion boundary condition, if a smaller amount of accreted energy  $\Ladd$ is assumed. This is an important result because it does not invalidate  previous studies based on the former treatment for hot accretion \citep{Baraffe-2009, Hosokawa-2011, Baraffe-2012}. 

The outcome of this work depends on various assumptions and uncertainties that we discuss below. The first question regards the impact of using 2D instead of 3D geometry. It is well known that 2D description of turbulent convection in stars gives different results compared to 3D simulations \citep[see e.g.][and references therein]{Meakin-2007}. Exploring the same range of parameters used in the present study with 3D simulations is currently unaffordable.  We do not, however, expect the qualitative results to be altered by the limitations inherent to the 2D description of convection.  

One of the most interesting results is the dependence of behaviour on the entropy of the accreted material and the subsequent formation of a hot surface layer.  We find no obvious argument why such layer would not form in 3D, though it could conceivably change the value of $\alpha$ at which this happens. Moving to 3D geometry will have an impact on the development of meridional flows at the surface of the accreting objects and how accreted material is redistributed over its surface. 
But given our assumption that the accreting area covers the entire surface of the star, inspired by the boundary layer simulations of \cite{Kley-1996} or by the idea of a spreading layer surrounding the whole accreting object,  the transition from 2D to 3D geometry should not change the general picture described in this paper.  

Extending the simulations to 3D could be relevant if the accreting surface area does not
cover the whole star, but only a fraction of it. This is an assumption that should be considered in future work, as it will change the outer boundary conditions and how the accreting star radiates its energy. 
Extending the simulations to 3D would also be important if
including the accretion of angular momentum and analysing its redistribution within the stellar interior. This problem has not been addressed, because our 2D simulations do not account for rotation of the object and angular momentum of the accreting material. Given our extremely poor knowledge regarding the rotational properties of embedded protostars and of the loss of angular momentum of material in a boundary layer, such a simple configuration may be realistic in some cases,
and at the very least a good baseline for comparison with potentially more sophisticated future models.

The treatment of the surface layers and of the boundary conditions, as well as the grid resolution, are other sources of uncertainty. In a parallel study, we are investigating the sensitivity of convection properties to the grid resolution and the boundary conditions in the same stellar model as used in this work (Pratt et al. 2016). We do expect quantitative results  such as the value of $\eacc$ that characterises the change in behaviour in the simulations and the formation of a hot surface layer, the surface temperature and the thickness of the high entropy surface layer of accreted material in the case of hot accretion to depend on the setup ({\it i.e.} treatment of surface layers, boundary conditions and/or grid resolution). But these setup effects may be covered by the large uncertainty contained in the free parameter $\alpha$ regarding the amount of accretion energy transferred onto the surface of the accreting object. 

One key assumption of our study though concerns the treatment of accretion through a simple accretion boundary. Further exploration of different types of boundaries to account for the accretion process are required to determine the impact of our assumption. Finally, present simulations are limited by the use of a fixed spatial grid that does not allow the stellar radius to vary. The 1D simulation including an accretion boundary condition indicates that after about 1\,yr, the envelope starts to expand and the radius increases. For this reason, it is incorrect to simulate  longer times with our multi-D simulations, 
unless some grid movement is allowed for, 
because our  multi-D framework does not account for the expansion of the star in the hot accretion case.  We are exploring the possibility of implementing a moving grid in MUSIC, which would be required not only for accretion studies but also for investigating stellar pulsation problems. 

This work is the  first attempt to describe the multi-dimensional structure of accreting young stars. Understanding the accretion process from first principles  with numerical simulations is a formidable physical and numerical challenge that we do not pretend to fully address with the present multi-dimensional simulations. Our motivation is to use multi-dimensional simulations to analyse the validity of  accretion treatments used in 1D stellar evolution studies and to improve such treatments for further studies and for the interpretation of observations. More broadly, our motivation is to perform similar multi-dimensional studies for other key processes  of stellar evolution such as convective overshooting, rotation, or pulsation. This pioneering work hopefully shows the potential of using time implicit multi-D simulations to improve our understanding of stellar evolution. 

\begin{acknowledgements}
I.B. would like to dedicate this work to Francesco Palla 
who recently passed away.
Part of this work was funded by the Royal Society Wolfson Merit award WM090065, the Consolidated STFC grant ST/J001627/1STFC, the French ÒProgramme National de Physique StellaireÓ (PNPS) and ÒProgramme National Hautes \'EnergiesÓ (PNHE), and by the European Research Council through grants ERC-AdG No. 320478-TOFU and ERC-AdG No. 341157-COCO2CASA. This work used the DiRAC Complexity system, operated by the University of Leicester IT Services, which forms part of the STFC DiRAC HPC Facility (www.dirac.ac.uk ). This equipment is funded by BIS National E-Infrastructure capital grant ST/K000373/1 and STFC DiRAC Operations grant ST/K0003259/1. DiRAC is part of the National E-Infrastructure. This work also used the University of Exeter Supercomputer, a DiRAC Facility jointly funded by STFC, the Large Facilities Capital Fund of BIS and the University of Exeter.
\end{acknowledgements}

\bibliographystyle{aa}
\bibliography{references}

\end{document}